\documentclass[manucript]{acmart}

\usepackage{amsmath}
\usepackage{graphicx}
\usepackage{booktabs}
\usepackage{float}
\usepackage{hyperref}
\usepackage{longtable}
\usepackage{lscape}
\usepackage{booktabs} 
\usepackage{geometry} 
\usepackage{array}    
\usepackage{courier}
\usepackage{pifont}
\usepackage{soul} 
\usepackage{xcolor}
\usepackage{xcolor}
\usepackage{courier}

\AtBeginDocument{%
  }

\copyrightyear{2025}
\acmYear{2025}
\acmConference[CHI '25]{CHI Conference on Human Factors in Computing
Systems}{April 26-May 1, 2025}{Yokohama, Japan}
\acmBooktitle{CHI Conference on Human Factors in Computing Systems (CHI
'25), April 26-May 1, 2025, Yokohama,
Japan}
\acmDOI{10.1145/3706598.3713171}
\acmISBN{979-8-4007-1394-1/25/04}

\begin{document}

\title[User Perceptions of Warning Label Designs for AI-generated Content on Social Media]{Labeling Synthetic Content: User Perceptions of Warning Label Designs for AI-generated Content on Social Media}

\author{Dilrukshi Gamage}
\authornote{All authors contributed equally to the research reported in this paper}
\orcid{0000-0002-6840-7718}
\affiliation{%
  \institution{University of Colombo School of Computing}
  \city{Colombo}
  \country{Sri Lanka}
}
\email{dilrukshi.gamage@gmail.com}

\author{Dilki Sewwandi}
\authornotemark[1]
\affiliation{%
  \institution{University of Colombo School of Computing}
  \city{Colombo }
  \country{Sri Lanka}
}
\email{dsewwandi2001@gmail.com}
\author{Min Zhang}
\authornotemark[1]
\affiliation{%
  \institution{The Open University}
  \city{Milton Keynes}
  \country{United Kingdom }
  }
\email{min.zhang@open.ac.uk}

\author{Arosha K. Bandara}
\authornotemark[1]
\affiliation{%
  \institution{The Open University}
  \city{Milton Keynes}
  \country{United Kingdom}
}
\email{arosha.bandara@open.ac.uk }


\begin{abstract}
  In this research, we explored the efficacy of various warning label designs for AI-generated content on social media platforms---e.g., \textit{deepfakes}. We devised and assessed ten distinct label design samples that varied across the dimensions of sentiment, color/iconography, positioning, and level of detail. Our experimental study involved 911 participants randomly assigned to these ten label designs and a control group evaluating social media content. We explored their perceptions relating to 1) Belief in the content being AI-generated, 2) Trust in the labels and 3) Social Media engagement perceptions of the content. The results demonstrate that the presence of labels had a significant effect on the user's belief that the content is AI-generated, deepfake, or edited by AI. However their trust in the label significantly varied based on the label design. Notably, having labels did not significantly change their engagement behaviors, such as 'like', comment, and sharing. However, there were significant differences in engagement based on content type: political and entertainment. This investigation contributes to the field of human-computer interaction by defining a design space for label implementation and providing empirical support for the strategic use of labels to mitigate the risks associated with synthetically generated media.

\end{abstract}

\ccsdesc[500]{Human-centered computing~User studies}
\ccsdesc[500]{Human-centered computing~Empirical studies in collaborative and social computing}

\keywords{Generative AI warnings, warning label design, user perceptions, deepfake, AI content label}

\received{Todate}
\received[revised]{15 Dec 2025}
\received[accepted]{17 January 2025}

\maketitle

\section{Introduction}
Advances in Artificial Intelligence (AI) technologies have greatly expanded the opportunities for content creation and sharing. Some are deepfakes or AI-generated content, such as images, audio, video, or text which are synthetically edited or created using algorithms that are easily accessible to a wide range of end users.  Since anyone can generate these sophisticated and highly realistic images, videos, or audio using generative AI technologies, the risk of shared mis/disinformation is rising exponentially. The majority of this content draws significant attention and engagement on social media platforms - resulting in their 'going viral' regardless of whether the content is fake or real. For example, AI-generated influencers share posts that are not necessarily false but they are not real people. In other cases, they cause harm such as creating and sharing synthetically edited images of a real person showing an action that never took place to mislead others on social media. Unlike the regular 'cheapfakes' (i.e. fake information that has been generated using manual editing techniques), AI-generated content has distressed the digital social media space, threatened democracy, and impacted society on many levels. 

\begin{figure*}
    \centering
    \includegraphics[width=1\linewidth]{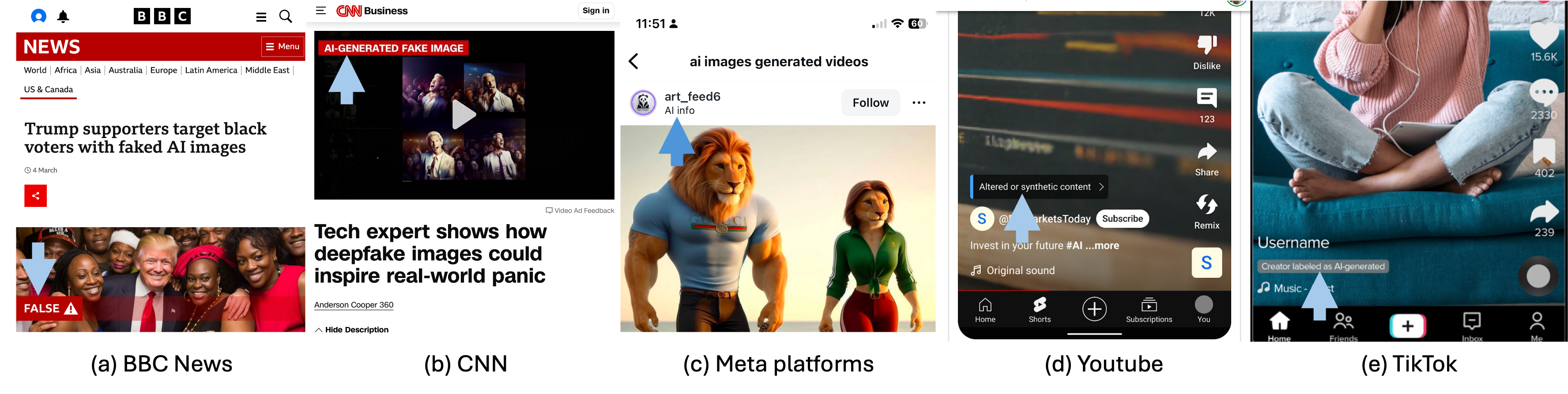}
    \caption{Examples design choices adopted for AI-generated content warning labels, highlighted by blue arrow annotation, on different platforms---BBC News, CNN, Meta Platforms (Instagram, Facebook), YouTube, and TikTok.}
    \label{fig:literature}
    \Description{ }
\end{figure*}

In an attempt to address these emerging challenges, regulatory actions were proposed by various legislative bodies. For example, the transparency obligation of the EU AI act imposed labeling the AI-generated content by developers and users of deepfake technologies \cite{labuz2024deep}.  Similarly in the United States, the White House executive order strengthened the laws relating to the use of AI and mandated that users need to be informed about AI-generated content \cite{worsdorfer2024biden}. The California AI Transparency Act (SB-942) became effective on January 1, 2025, and aims to enhance transparency around AI-generated content, including images, audio, and video. Targeting AI providers with over one million monthly users, the law mandates several disclosure measures: e.g., providers must offer users a visible disclosure option to label content as AI-generated, incorporate imperceptible disclosures (such as digital watermarks) indicating content provenance, and provide a free, publicly accessible tool to verify if the content was created or modified by their AI systems. Currently, we are seeing a variety of methods being used to indicate content being deepfake or AI-generated by news organizations. At the same time, major platforms have been attempting to adhere to the transparency needs by the regulatory bodies by articulating a variety of methods (see Figure \ref{fig:literature}). For example, adding markup labels to AI content (Google), or adding a new label with the wording "AI Info" to video, audio, and image content (Meta). Despite this variety of labels and mechanisms to indicate deepfake or AI-generated content, there is limited empirical evidence of how user perception and engagement behaviors are affected by the presence of such labels.

Previous studies showed how misinformation ``warning labels'' can reduce people’s likelihood of believing such content \cite{jia2022understanding} and may subsequently reduce liking, commenting \cite{lim2023effects}  and sharing intentions\cite{mena2020cleaning} for false and unsubstantiated claims. Researchers draw on these results to suggest that labeling AI-generated content may lead to behavioral effects similar to those of misinformation labeling. However, they also highlight the need for empirical research to understand the nuances of how these labels should be implemented and applied in the context of AI-generated content \cite{wittenberg2024labeling}. Some preliminary work explores potential terms or words that could be used in labels to inform users about AI-generated content and recommends that designers of deepfake warning labels need to determine and understand the objective of the label \cite{epstein2023label}---i.e. whether it is for the purpose of informing users about AI-generated content or misleading content. Such studies provide statistical evidence of embedded wording that could be incorporated to label AI-generated content. However, the nuances of designing a content warning label to meet the transparency regulations and its applicability, understandability, and adaptability of such labels and their effects on human behavior need careful exploration. The design of warning labels needs to consider many dimensions such as color, iconography, positioning, and level of detail used to inform users about AI-generated content. Therefore, we position our research in two major ways---First, we explored the design space for warning labels based on a review of the algorithmic misinformation detection literature and label implementations being proposed by different online platforms. Second, we conducted an experiment to investigate the efficacy of example labels derived from the design space.

To position our research we found that existing misinformation labels and efficacy studies~\cite{martel2023misinformation} lack the underpinning investigation of the perception of deepfakes and AI content warning labels to behaviors and understanding of the design space that must be considered when labeling AI-generated content. Specifically, how users will interact with these new types of labels in social networks remains under-explored in the field of human-computer interaction. Early research work on labeling AI-generated content suggests considering factors such as the intended audience, message, goals, process, and platform of the content creator when designing a labeling system for authenticity \cite{burrus2024unmasking} and recommends being mindful of terminology without assuming that the average person knows exactly what generative AI entails\cite{burrus2024unmasking}. However, it is important to consider other aspects of the design for warning labels for AI-generated content, which leads us to ask: \\
\\
    \fbox{
    \begin{minipage}{\dimexpr\linewidth-2\fboxsep-2\fboxrule\relax}
    \centering
    \textbf{RQ1:} What are the dimensions of the design space for warning labels for AI-generated content?
    \end{minipage}}
\\


However, the design space for warning labels for AI-generated content is multi-faceted, needing iterative refinement and evaluation of candidate designs to assess their real-world impact. However, to perform some initial empirical evaluations of label efficacy, we derive 10 sample warning label designs out of the space and explore user perceptions to validate if the designs are effective. We operationalize this by incorporating measurements---1) Perceived believability of the content, and 2) Perceived trust of the warning label. These measurements provide evidence of how users trust the labels they see on the social media platform and thus their belief in the content tagged by the label. For this aspect of our study, we specified the following research questions:\\
\\
    \fbox{
    \begin{minipage}{\dimexpr\linewidth-2\fboxsep-2\fboxrule\relax}
    \centering
    \textbf{RQ2:} How do warning labels affect users' belief of whether the social media content is AI-generated?\\
    \textbf{RQ3:} How does the design of the warning label affect users' trust in the warning? 
    \end{minipage}}
\\

Warning labels designed for AI-generated content enable transparency, but it is not known if this transparency leads to trust in the platform that displays the warning labels. Additionally, the signals provided by warning labels may affect the users' engagement behaviors with the content content. By 'engagement' we refer to the common interactions with content on social media platforms such as "Like" reactions, commenting, and sharing content. This leads us to investigate:\\
\\
    \fbox{
    \begin{minipage}{\dimexpr\linewidth-2\fboxsep-2\fboxrule\relax}
    \centering
    \textbf{RQ4:} Is there a relationship between trust in the warning label and trust of the associated social media platform? \\
    \textbf{RQ5:} Do warning labels on AI-generated content affect users' engagement behaviors? 
    \end{minipage}}
\\


Incorporating literature and the current context of AI-generated content such as deepfakes, we derived a design space with four key dimensions---(1) label sentiment, (2) icon and color of the label, (3) position of the label, and (4) level of detail. We empirically tested 10 different sample labels derived from this design space. Our evaluations used 911 participants and results indicated that in general, users' trust varied depending on the label designs. We also found that users significantly believed that the content was AI-generated as a consequence of trusting the labels. However, having warning labels did not have a significant effect on engagement behaviors such as sharing, commenting, and like reactions. Overall, we outline the effects and implications of having warning labels for AI-generated images, to make two specific contributions: 
\begin{enumerate}
    \item  A design space for warning labels for AI-generated content, providing a framework that online platform designers and researchers can expand and experiment with.
    \item Empirical evidence of the efficacy for 10 prototypes derived from this design space, highlighting design implications for designers, developers, and policymakers working on warning labels for AI-generated content.
\end{enumerate}

\section{Related Work}
Our work builds on prior research in the mis/disinformation domain which investigates ways to mitigate the spread of false or misleading content. Specifically, under the misinformation interaction cycle, we focus on automated warning label strategies at the intersection between emerging industrial strategies intersecting and regulatory requirements. In this section, we provide background on current interventions being adopted to label deepfakes and AI manipulation of content on social media platforms. We also summarize the status of the regulatory actions and emerging standards for deepfake warnings alongside the previous research on automatic misinformation warning labels. 

\subsection{Social Media Platform and Regulatory Responses to Deepfakes}

In this research, it is crucial to explore the process and design discussions considered by the major social media platforms on the deepfake content that is shared on their platforms. For example, Google recently announced that it is making an effort to mitigate the impact created by deepfakes---e.g., by preventing deepfakes from appearing in search results, enabling reporting of fake explicit imagery, as well as the removal of these images from Google platforms \cite{MITReview@2024}. At the same time, Meta released a statement of intent to work with industry partners on common technical standards for identifying AI content, including video and audio. Although these statements are constantly evolving, at the time of writing we found that Meta platforms like Facebook, Instagram, and Threads are using labels to indicate if any content is created by AI \cite{Meta}. TikTok claims they are the first video-sharing platform to implement Content Credentials, which is an open technical standard that provides publishers, creators, and consumers with the ability to trace the origin of different types of media \cite{CCredential}. This technology will embed a digital watermark in the AI content, which identifies details such as when, where, and by whom the content was created. TikTok announced that they will use such data to flag the content as being generated or manipulated using AI technologies. Before that, they have also used labels to indicate AI content. In March 2024 Google announced how YouTube will introduce an upload flow requiring creators to share when the content they’re uploading is meaningfully altered or synthetically generated and seems realistic \cite{Youtube}. Some of the sample designs for these warnings, taken from websites and announcements made by the different platform service provides, are shown in Figure \ref{fig:literature}. Overall, this shows that major platform providers want to use warning labels to help their users identify AI-generated content, but there is limited evidence on the efficacy of these label designs. 

\subsection{Designing for Mis/Disinformation}
Even though its use is emerging, we are witnessing that deepfakes are posing a threat to democracy by making it easier for people to create false information. While the challenges associated with mis/disinformation are not new to social media, the rapid proliferation of generative AI technologies have enhanced the ability of online actors to influence public opinion, polarize societies, and undermine trust in institutions. Misinformation refers to the spread of false or inaccurate information, regardless of intent, while disinformation specifically involves the deliberate creation and dissemination of false information to deceive or manipulate. The growth of social media and digital platforms has amplified the reach and speed at which such content spreads, making it a critical issue for researchers, policymakers, and technologists. To combat the spread of misinformation and disinformation, various research interventions have been proposed and implemented. These interventions range from educational initiatives aimed at improving digital literacy to technological solutions such as AI-driven fact-checking tools. To better understand the design space for these interventions, researchers provide a misinformation engagement framework that maps and systematizes the core stages that people engage with online misinformation \cite{geers2023online}. 

This explains interventions can be designed when 1) Users select the source,  2) Information selection, 3) Evaluation, and 4) Reaction stages. At stage one, commonly used interventions include source credibility labels  \cite{otis2024effects}, and friction  \cite{ershov2024sharing}. When users scroll through a social media news feed, read a headline, clicking on an article, they are in stage two of the framework, and the interventions proposed aim to help make critical judgments easier and more effective. These interventions mostly have different labels and warning signs which help users to make informed decisions \cite{martel2023misinformation}. Similarly, some other interventions aim to help users master \textit{`critical ignoring'} which is a recommended habitual behavior that could be taught as part of the school curriculum \cite{kozyreva2023critical}. When it comes to evaluating the accuracy of the information and/or credibility of the source, previous research indicates a wide variety of interventions, including approaches like 'debunking' and fact-checking \cite{lewandowsky2020debunking}.

Researchers refer to the stage where users attempt to engage with misinformation as the  \textit{`reaction stage'}, involving the judgment of whether and how to engage with the information (e.g., by commenting, `liking', or sharing). Interventions such as accuracy prompts \cite{pennycook2021shifting} , friction \cite{fazio2020pausing}, and cultivating social norms \cite{andi2021nudging} are proposed to reduce the engagement at this stage.  A systematic review of misinformation highlighted a taxonomy that depicts that the interventions can be identified by their design, users' interaction, and the timing of the intervention \cite{hartwig2024landscape}. The design of misinformation interventions is categorized into warning labels \cite{bhuiyan2018feedreflect};  correction/debunking activities, which also can be displayed in naturally occurring or artificially generated user comments; comments from officials that correct misinformation or with links to fact-checking websites \cite{bode2015related}; or expert sources \cite{bhuiyan2021designing}. On the other hand, some interventions require active interactions where the interventions require users to actively interact with a countermeasure such as a link to confirm before sharing; overlays on Facebook and X (formerly Twitter) posts; and pop-ups \cite{jahanbakhsh2022leveraging}. More commonly, the designs for interventions are focused on the timing of the encounter, such as interventions focused on Pre-exposure ---accuracy nudge before a news-sharing task, or `pro-truth pledge' to engage in more prosocial behavior; During exposure---interventions designed such as algorithmic corrections next to a post, user comments underneath a post, or warnings; or Post-exposure---where interventions designed as warnings, responses by health authorities, debunking text including plausible scientific explanations to close gaps in mental models or it could be at the time of sharing or on request of the user \cite{hawa2021combating}. 

All of these intervention designs provide insights to combat misinformation, yet the real issue lies when the design needs to scale and have high efficacy. Deepfakes are spreading too fast and users find it almost impossible to distinguish them from real content. It is essential to use technologies to detect deepfakes quickly, but this information also needs to be communicated to the user in a trustworthy manner to reduce engagement behaviors that lead to harmful consequences. Labeling the content is much discussed in the GenAI literature, but the efficacy of different label designs is not empirically validated. Therefore, in this research, we focus more on label design and test for impact to understand the scaling effect.  

\subsection {Misinformation Labels and Effect}
A widely adopted method of countering misinformation is having \textit{``labels"} to communicate an informed warning to users based on fact checkers or system decisions \cite{martel2023misinformation}. These labels may be designed directly on or alongside the content. Many social media platforms have different label designs---Facebook \cite{ohlheiser2016facebook}, Instagram  \cite{oeldorf2020ineffectiveness}, and Twitter  \cite{papakyriakopoulos2022impact} have some fact-checkers taglines or community taglines as labels. As we found for the mis/disinformation intervention designs, the warning labels differ from related informational interventions such as pre-bunks or corrections \cite{chan2017debunking} which are delivered before or after exposure, respectively. One of the prominent tasks of the fake news identification process is how the results of detection should be communicated to the user. Researchers argue its not only the method of delivering the message but also the method used in identifying the mis/disinformation---i.e. credibility also matters. The trust for computation-based detection and prevention of fake news and decision-aid methods (such as using fact-checkers opposed to warnings by ML methods) to warn users when a piece of fake news has been explored \cite{seo2019trust}. Researchers also identify that interventions can be just an indicator of contextual information \cite{guo2023seeing}, including the use of different colors for the label  \cite{nassetta2020state}; relevant words for misinformation classification \cite{lim2023effects}; generic tips to detect misinformation; or infographics \cite{domgaard2021combating}. This highlights the importance of research to systematically study the effects of different label design dimensions, e.g.,  their positions, colors, or fonts \cite{gao2018label}. 

Notably, all of these interventions targeted traditional misinformation, but exploration of intervention design for emerging types of misinformation, such as those using AI-generated content, is significantly limited. Recent ongoing research work examined that if given the warning of the deepfake content, the likelihood of correctly detecting deepfake videos is approximately twice as likely as the inauthentic video in the control group \cite{lewis2022content}. Similarly, it was shown that the text in the label affects users' ability to detect if the content is a deepfake or not, depending on the intended goal of the label \cite{epstein2023label}. In this study,  participants found that the term "AI-generated" was most consistently associated with content generated using AI, but if the goal is to label misleading content, the term "AI-generated" performed poorly. As a solution to this, the terms “Manipulated” and “Not Real” were found to be the most consistent for misleading information. With such combinations, it is clear that the design space for labeling is wide and there is a need for robust evidence to identify effective labels for AI-generated content.

Some recent studies have addressed this gap, providing evidence demonstrating the utility of warning labels, by substantially reducing people’s belief in and sharing of AI-generated content \cite{wittenberg2024labeling}. However, rather than following the approach taken by traditional mis/disinformation labeling, this work highlights the importance of establishing the objectives of the labeling for AI-generated content. Currently, policy and regulation are between the use of labeling to communicate to viewers the 'process' by which a given piece of content was created or edited (i.e. with or without the use of generative AI). The policy expectation is to inform the viewer; the expectation is to reduce the likelihood of sharing or mitigating the risk of misleading or deceiving users. However, empirical evidence for these results is limited. Social behaviors associated with AI-generated content are still emerging, and further research is needed to highlight the issues and challenges that must be considered when designing, evaluating, and implementing labeling policies and programs. Specifically, these investigations need to (1) determine the types of content to label and how to reliably identify this content at scale, (2) consider the inferences viewers will draw about labeled and unlabeled content, and (3) evaluate the efficacy of labeling approaches across contexts \cite{wittenberg2024labeling}. Contributing to this empirical work, our research provides a design space in which labels can be designed to warn users about AI-generated content. We then conduct experiments to explore the features of warning labels that may enhance their efficacy in understanding user behavior and reducing misleading deepfake content. In the next section, we explain the methods we used to derive the design space for warning labels for AI-generated content and empirically evaluate the efficacy of different warning labels derived from this design space.

\section{Methods}
Our research adopted a mixed-method approach, first using the qualitative investigation of real-world AI-generated content warning label designs and related literature to identify a design space, followed by a combination of quantitative and qualitative techniques to evaluate candidate label designs derived from the design space.

\subsection{Design Space for AI-generated Content Warnings}

Warning labels for false or misinformation are not new; however, the challenge is designing warning labels that are specific to highlighting AI-generated content and understanding users' responses to such labels. We started exploring design specifications based on an evolving body of misinformation research. We found some specific coalitions working together to provide transparency and authenticity to the media---C2PA \cite{C2PA} which provide Content Credentials (CR) \cite{CCredential} standards and Content Authenticity Initiative\cite{contentauthenti}, Partnership AI \cite{partnership} that jointly works with C2PA to bring interventions to combat the authenticity issues of AI-generated content. Some techniques move further in proving correction strategies as provenance using cryptographic techniques which allows media authoring tools to embed cryptographically signed provenance information into media metadata. Visualization of provenance is one direction for AI-generated content, but previous warning labels provide a rich background on design directions based on user perceptions for effective warning labels for communicating and improving transparency. Combined with misinformation research, this work provides insight into the cognitive processes that influence the way individuals assess the credibility of information. We also incorporated principles of human-computer interaction (HCI) which emphasizes the importance of user-centered design. This helps to logically connect a framework's attributing design dimensions to the design space. Such a framework provides a ground where we can design combinations of attributes in warning labels and evaluate them by focusing on how labels are perceived and interacted with by users.

In addition to these academic insights, we also explored how some key platforms are changing their designs related to labeling deepfake or AI-generated content. Since the associated regulatory requirements for AI content labeling are still evolving, we witness constant changes in how platforms inform users about their practice, i.e. Meta's labeling approach updates~\cite{Metachange}. Four researchers who are experts in mis/disinformation and user-centered design articulated the initial dimensions of the design space. Later, through several debriefs, brainstorming sessions, and mapping to requirements of current policies and scholarship, the team outlined and expanded the subcategories. Based on the goals, we made options that the framework not only captures the user's attention but also promotes cognitive engagement, fostering a critical assessment of content authenticity. Integrating these elements involves a careful balance to maximize both the informative and the cautionary functions of the labels. 
Our approach combines expert brainstorming, prior research results, and gray literature from platform providers, to identify the dimensions and options to include in the design space. Labels derived from the design space must inform users so they can make informed decisions and change behaviors. We believe our design space provides an initial framework that can be refined through iterative testing to optimize the efficacy of warning labels for AI-generated content in different contexts. In this paper, we present a first iteration where we provide insights into our experiments that tested our design options in controlled settings. Our approach ensures that the labels are adapted to the specifics of different content categories (e.g., political vs. entertainment) and diverse user demographics, who may vary significantly in their digital literacy and familiarity with generative AI technology.

\subsection{Deriving Design Samples}
Our initial framework categorizes the elements of a label into four main dimensions: sentiment, iconography/color, position, and level of detail, each selected based on empirical evidence and considering the current practices applied to some major social media platforms and news organizations. Once we categorized the dimensions, subsequent brainstorming sessions led to identifying and creating artifacts (warning labels) based on the design landscape with the aim of evaluating their efficacy. One challenge we faced during the design of the sample prototype warning labels is that the design space is broad, with a range of design options, leading to a complex evaluation process. In the absence of a systematic evaluation of all warning label designs that could be derived from the design space, the choice of any given design will be subjective.

Therefore, before mapping the design space for warning labels for AI-generated content, as a team we tried to objectively understand the goals of such labels and their current context. In the context of mis/disinformation, the primary goal was to distinguish the fake from the real content. However, in the context of AI-generated content, we understand that it has a new context where not all the content created using AI is fake---i.e., enhancing an image using AI or sometimes there is a new context/person created that never existed. Facebook specifically outlined its criteria for labeling (and not labeling) content. The label had the wording "AI info" and does not specify if the content is edited or newly created~\cite{Metalabel}. In our designs, we prioritize the importance of communicating with the user about the use of AI before they engage with the content on social networks. However, narrowing down the design space to 10 sample designs to evaluate was mainly based on the current demand reflected in regulatory acts such as the EU AI act\cite{EuAIact}, guidelines and frameworks to govern the AI sector on a federal level such as executive orders promoting the use of trustworthy AI in the Federal Government Act in the US, and other AI policies being adopted around the world. We set our primary objectives in designing samples from the design space based on the fact that labels will communicate the \textit{Content Creation Process} (indicate whether the content was generated or altered using AI tools); and reduce \textit{Misleading Effects} (reduce likelihood that content deceives users, regardless of its creation method).

\subsection{Evaluating Design Space for Warning Labels Using Prototypes} 

To understand the efficacy of warning labels designed for AI-generated content, we created 10 prototype labels based on our design space. Then we added these warning label prototypes to different social media image posts. By doing that, we aim to understand various aspects relating to such labels---1) users' engagement behavior in social media in the presence of such labels, 2) their trust in such labels and their trust in the social media platform if they see such labels, 3) users' belief of whether the content is a deepfake / AI-generated or significantly altered by AI, 4) whether having labels made user think before they engage in social media, and 5) whether having labels helped them to take an informed decision. With all of these, we specifically examined which label design provides the most representative efficacy. All our data, Python code, and Qualtrics Survey files are provided in the Supplementary folder and subsequently available in the Open Science Framework (OSF) Link - \href{https://osf.io/m8sg2/}{https://osf.io/m8sg2/} for greater reproducibility of the research.

\subsubsection{Experiment Design Using Warning Label Prototypes in Social Media}
To test the efficacy and related aspects of sample label designs, we conducted a control group and 10 treatment group experiments. All groups evaluated eight images taken from social media where 4 images were deepfakes, AI-generated, or edited, and 4 images were real. We designed the image with a frame to mimic a typical Facebook or Instagram-like post in the feed where it has a "like" reaction, a Comment icon, and a Share icon. Each image represents either a political or entertainment type of content balanced 50-50 for each group of samples. The 8 images (2 Deepfake, 2 Real Political images, and 2 Deepfake, 2 Real Entertainment pictures) depicted human subjects. The control group of participants were shown these 8 images without any label embedded in the images. Each treatment sample group uniquely sees only one type of label in the eight images. Since we had 10 labels designed, each label was added to 8 images in each group creating 10 treatment groups seeing the same images but with different labels. We include the eight detailed images used in the experiment in the Appendix \ref{sec:Image sources for the experiment}.

\subsubsection{Data Collection and Measurements}
All participants received a pre-survey, treatment, or control question and a post-survey authored using the Qualtrics survey platform. The pre-survey collected participants' social media and internet habits to understand general behaviors across demographics where we can trace if there is an interaction to the effects caused by our intervention---i.e. different warning label designs. We asked six questions where we collected: (Pre-\(Q_1\)) Internet browsing behavior, (Pre-\(Q_2\)) social media browsing hours, (Pre-\(Q_3\)) Names of social media platforms used regularly, (Pre-\(Q_4\)) likelihood of using the social media platform to type messages, upload content such as videos, photos, or voice music on social media, (Pre-\(Q_5\)) likelihood to engage (like, comment, other reactions) with such content posted by others on social media, (Pre-\(Q_6\)) Likelihood to share such content posted by others on social media. Except for Pre-\(Q_3\), which has check boxes for social media platform names, all other questions contain Likert scales from 1 to 5, indicating that the lower likelihood number is for a lower score and the higher number is for a higher score.

In the experiment phase, for both the control and treatment groups, we measured (\(Q_1\)) Familiarity to what they see in the image, (\(Q_2\)) If they recognize the people in the image \textit{\textsc{[Yes, Unsure, No]}} (\(Q_3\)) their belief if the image is AI-generated or AI edited significantly, their likelihood to ``Like'' (\(Q_4\)), Comment (\(Q_5\)), and Share (\(Q_6\)) that post on their social media feed, and finally (\(Q_7\))  if they think the image is fake or real \textit{\textsc{[Fake, Real]}}. Except for (\(Q_2\)) and (\(Q_7\)), all questions were measured using the Likert scale 1 to 5 measurements, i.e. for (\(Q_1\))- \textit{\textsc{[Extremely unfamiliar, Unfamiliar, Neither familiar or unfamiliar, Familiar, Extremely familiar]}}.

For each treatment group (each seeing a different sample warning label), we also measured their (\(Q_8\)) trust in the label, (\(Q_9\)) trust in the social media platform if they see that label, and (\(Q_{10}\)) their likelihood of engaging (through a like, comment, share) with the content if they see the label, (\(Q_{11}\)) whether the label helped them to take informed decision and finally (\(Q_{12}\)) who would be most suited to create and tag the image with the label they saw. While all others had a 1 to 5 Likert scale in the measurements, responses to (\(Q_{12}\)) allowed multiple selection from the options:\textit{\textsc{[-The social media platform itself}, \textsc{-The tool that generated this image}, \textsc{-It should be tagged by Fact checkers},\textsc{-It should be tagged by the social media users all together},\textsc{-It should be tagged by the user who upload the AI images]}}.

Out of these, we derived measurements for our main dependent variables: \textsc{Trust-In-Label}, \textsc{Belief-Of-Content}, \textsc{Trust-In-Platform}, \textsc{User-Engagements (Like | Comment | Share)}, \textsc{User-Informed-Decisions}, \textsc{User-Familiarity}. We decided not to use any composite measurements, but simple direct measurements. For example, ``Trust'' as a concept itself can be measured in multi-item verified scales~\cite{prochazka2019measure}. We specifically did not want the participants to evaluate the whole content with the label as the content can be true of false that a user may or may not know. Our intention was to distinguish our measured observations on whether participants trusted the label we designed regardless of the content. We found that direct measurements of Likert scales on warning labels trust were adopted in many repeated similar studies with social media intervention~\cite{pennycook2020implied}. Similarly, we see that the believability of content and sharing intentions were directly measured using Likert scales~\cite{pennycook2021shifting,pennycook2022nudging}.

The independent variables of our study represent the 10 warning label design samples which we list as; \textsc{Sample-1} to \textsc{Sample-10}. These samples were designed based on the design space which has many dimensional choices. Rather than treating each as a dimensional factorization that can be analyzed as a binary present/absent category, we use each design as a separate treatment as we intend to rapidly explore which designs may have possibilities to implement as warning labels. In addition to the design samples, we also used \textsc{Content-Category} as an independent variable to understand whether political or entertainment-type content has an impact on user perceptions measured by dependent variables.

In the post-survey, participants' demographics were collected along with their familiarity with creating and sharing deepfakes. Although we had the location of the participants, we did not collect any details about participant's education and employment details. This is mainly because the intention was to investigate the perception of any regular Internet and social media user regardless of their education level of employment status. In detail, we collected the following data from the post-survey: (Post-\(Q_1\)) Age group [Age group ranges in 5 scales], (Post-\(Q_2\)) Pronouns [He, she, them, other], (Post-\(Q_3\)) if they have encountered any deepfakes or not [a text input], (Post-\(Q_4\)) whether users know how to create deepfakes or AI-generated images {yes, no], (Post-\(Q_5\)) if they have created themselves [yes, no], (Post-\(Q_6\)) if they think deepfakes can be created easily [yes, no], (Post-\(Q_7\)) How likely they are to upload deepfake or AI-generated content on their social media platform [Likert scale 1 to 5], (Post-\(Q_8\)) How likely they are to share deepfake or AI-generated content that they have seen on social media feed [Likert scale 1 to 5], and finally an open question to share a textual explanation of how they feel about the label indicating that the image is AI-generated and to provide suggestions on how the labels could be improved. Full Python code for the data analysis can be found in \href{https://colab.research.google.com/drive/10FupI7NsR7X7jiGzyEYkEEi1gMW0Bupf?usp=sharing}{Google Collab file}.

\subsubsection{Qualitative Measures}
The answers in (\(Q_{12}\)) provide multiple options for participants to indicate who could label content on social media platforms. During the data analysis phase, we discussed participants' responses in conjunction with the answers they provided in the post-survey questions to identify themes. Specifically, participants explained in text how they had previously encountered fake or AI-generated content and made suggestions to improve warning label designs.

\subsubsection{Procedure}
The experiment was approved by our Institution's Ethics Review Board. The participants were recruited through the Prolific platform and each participant was compensated at the recommended rate of the platform of \$12.00 USD per hour. The data collection was conducted in August 2024. The average completion time for our study was approximately 20 minutes per participant. We conducted a priori power analysis using G*Power version 3.1 to determine the minimum sample size required to test the significance of the tests. It indicated that the sample size required to achieve the 80\% power to detect a medium effect, with a significance criterion of $\alpha$ = .05, was N = 88 for repeated ANOVA measures between and within interactions with 11 groups. The same test conditions for ANOVA fixed effect special main effects provided a total sample size (N = 269). We conducted the study by collecting 1000 responses, to maximize the likelihood of exceeding the required sample size and ensure that each treatment group would be representative of the wider population.

When each participant completed the pre-survey, they were randomly assigned to the control group condition or one of the treatment conditions before they completed the post-survey. Figure  \ref{fig:flowchart} explains the study procedure, together with the experiment questions given to each control condition and the treatment condition. 

\begin{figure*}[ht]
    \centering
    \includegraphics[width=1.0\linewidth]{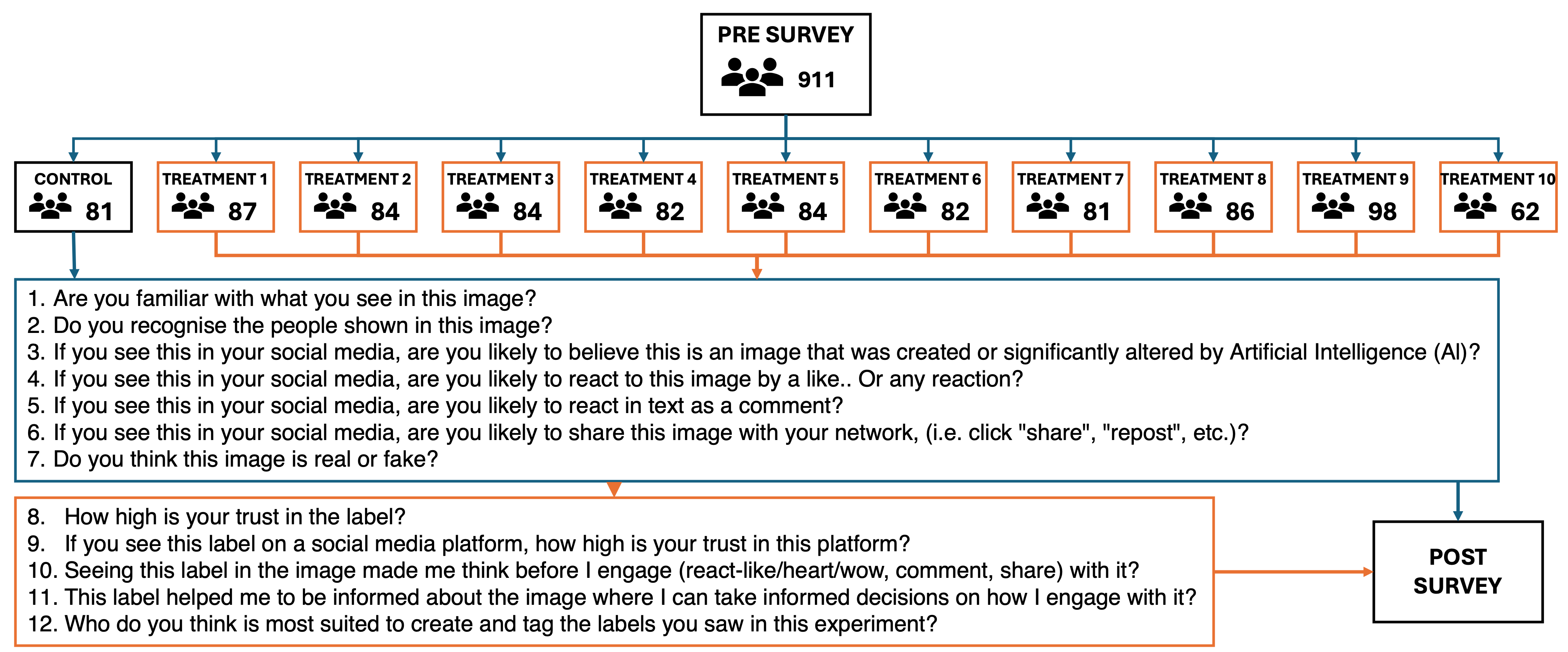}
    \caption{Experiment Procedure: 911 samples qualified, all of them took the same pre-survey, then randomly assigned to one of the treatment conditions, each using a different label design}
    \label{fig:flowchart}
    \Description{ }
\end{figure*}

\subsubsection{Data validation}
To make sure our participants are attentive to our questions, we placed an attention check question in our survey. The question checked whether participants took note of the label content and allowed us to eliminate responses from those who answered incorrectly. We also eliminated responses from participants who did not complete the survey. Finally, we performed a randomization check to verify that the allocation of participants to the different conditions had a similar distribution by age and gender.

\subsubsection{Participants}
Based on the inclusion checks, a total sample of \( n=911 \) was qualified after completing the survey. Among them, 248 participants were between 51--61 years, comprising 27.16\% of the total sample. This was followed by the 29--39 years age group, which includes 179 participants (19.61\%), and the 40--50 years age group, with 170 participants (18.62\%). The younger age group of 18--28 years accounts for 151 participants, or 16.54\% of the sample, while those aged 62--72 years make up 15.33\% with 140 participants. A small portion of the sample, 23 participants (2.52\%), identified as belonging to an ``Other age'' category. In terms of gender pronouns, the majority of participants identified with ``She/Her,'' representing 459 participants or 50.38\% of the total sample. This indicates that half of the respondents are women or individuals who identify with these pronouns. Meanwhile, 418 participants (45.88\%) identified with ``He/Him,'' showing a nearly equal representation of men. A smaller group of 30 participants (3.29\%) preferred not to disclose their gender pronouns, and only 2 participants (0.22\%) identified with ``They/Them,'' reflecting a very small representation of non-binary or gender non-conforming individuals in the survey. The social media usage data shows that the largest proportion of participants, 38.20\%, reported spending 1--2 hours per day on social media, which corresponds to 248 individuals. A notable portion, 24.04\%, or 219 participants, indicated using social media for less than 1 hour each day. A smaller group, comprising 20.09\% of the sample, or 183 participants, reported spending 2--3 hours per day on social media. Additionally, 10.76\% of participants, or 98 individuals, reported using social media for 3--4 hours daily. Only 6.92\% of participants, which equals 63 individuals, indicated spending more than 5 hours per day on social media. These results suggest that most participants engage with social media in moderation, with a majority using it for 1--2 hours each day. Analysis showed similar results for the Internet browsing behavior. See Appendix \ref{sec: Demographicdata} for details. 

\section{Results} \label{Results}
As explained, we had two primary objectives in this research---creating a design space for warning labels specifically for AI-generated content and evaluating the efficacy of the created warning labels. Our design space focuses mainly on four design dimensions: Label Sentiments, Icon/Color, Position, and Label Detail. From the design space, we derived 10 prototype warning labels. We first provide details of the design space and the derived prototypes. Next, we provide the results of the evaluations we performed using the measurements collected through our experiment.

\subsection{RQ1: What are the Dimensions of the Design Space for Warning Labels for AI-generated Content?}.
The design space for warning labels is very divergent, with many dimensions that may not easily be logically distinguished from each other. Our brainstorming sessions resulted in exploring fundamental categories for warning label designs and evaluating them. Because we are designing warning labels for AI-generated content, and upon agreement with the team, we identified four design dimensions and associated design options for these dimensions.

\subsubsection*{\textbf{Label Sentiment (LS):}} The tone and words in the warning can affect the perception of the warning label by the user. Specially the AI-generated content shared in platforms needs a warning label that should quickly attract the user's attention and signal what it is about. Specifically, it must provide language that informs about any hazards, and promotes or demotes any consequences confronting such content.
\begin{itemize}
    \item \textbf{LS1: Hazardous}(e.g., "Deepfake content, Danger, Fake" ): This option will inform the user the content is not real using judgmental language that highlights the hazard. It may affect the user's perception which may lead users to be cautious before interacting with the content.
    
    \item \textbf{LS2: Neutral - no mention of AI} (e.g., "Synthetically generated"): AI-generated content can be identified as synthetic content, which leaves open any judgment of it being benign or harmful. 
    
    \item \textbf{LS3: Neutral - AI mentioned} (e.g., "Made with AI", " AI info" ): Clearly indicates AI involvement, balancing informativeness with neutrality, suitable for an audience that might appreciate transparency about content generation without additional emotional coloring. By mentioning AI, it signals to the user that this is generated through AI providing contextual perceptions before judgment of any content
\end{itemize}

\begin{table*}
\begin{center}
    \caption{Summary of the label design framework for AI-generated content}
    \label{tab:frameworktable}
\begin{tabular}{cll}
\hline
\textbf{Design Dimension} & \textbf{Design Option} & \textbf{Examples} \\
\hline
Label Sentiment (LS) & LS1: Hazardous tone & BBC, CNN \\
& LS2: Neutral tone - not mentioning AI & Youtube, Adobe Content Credentials (CR) \\
& LS3: Neutral tone - mentioning AI & Meta (Instagram), ABC, TikTok \\
\hline
Icon/Colour (IC) & IC1: Warning: <!> / Red & BBC, CNN, ABC \\
& IC2: Neutral icon: [\includegraphics[width=0.3cm]{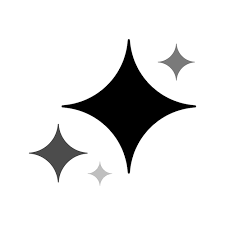}] / monochrome & Meta (Instagram), Canva Magic Studio, Gemini  \\
& IC3: No icon / monochrome & TikTok, Adobe CR \\

\hline
Position (P) & P1: On content (obscuring) & BBC, CNN, ABC \\
& P2: Above content & Meta (Instagram) \\
& P3: On content (non-obscuring) & TikTok, CNN, Adobe CR \\
\hline
Label Detail (D) & D1: Simple warning text & BBC, Meta, CNN, etc. \\
& D2: Detailed information & Adobe CR \\
\hline
\end{tabular}
\end{center}
\end{table*}

\subsubsection*{\textbf{Iconography and Color (IC):}} Visual cues are crucial in the rapid processing of information. The icons and color choices serve as immediate indicators of the nature and urgency of the content. The choice of icon style, warning or neutral, strives to balance visibility with user experience, avoiding alarm fatigue while maintaining effectiveness. Similarly, some colors are more indicative of danger (e.g., red and yellow).  This approach aligns with current practices seen in the platforms where icons and colors are used to quickly convey information status, such as in software applications and digital advertising.

\begin{itemize}
    \item \textbf{IC1: Warning icon/Red} (e.g., "Warning: <!>"): Using universal warning symbols and color to attract attention and denote caution, which prompts users to reconsider their trust in the content.
    \item \textbf{IC2: Neutral icon/Monochrome} (e.g., "Neutral icon: [\includegraphics[width=0.3cm]{aiicon.png}], combinations of diamond shapes used in many platforms to perform AI generating actions"): Offers a less intrusive alert that blends more seamlessly into the design of the platform, potentially reducing impact but maintaining aesthetic consistency. This icon could be something that the majority of platforms used to indicate AI generation in a manner that is intuitive to users.
    \item \textbf{IC3: No icon/Monochrome}: This minimalist option might be used where the platform desires the label to be as unobtrusive as possible, relying solely on text to convey the message.
\end{itemize}

\subsubsection*{\textbf{Positioning (P):}} The placement of the label affects its visibility and the potential to obstruct or enhance the content. Strategic positioning choices, such as overlaying content or placing labels around it, are informed by user interaction studies and regulatory guidelines that aim to ensure content visibility without significant interference.

\begin{itemize}
    \item \textbf{P1: On content (obscuring)}: Ensures visibility by placing the label directly on the content, though it may obstruct part of the image, which could affect user experience.
    \item \textbf{P2: Above content}: This placement keeps the label in close proximity to the content without obscuring any part of the image, balancing visibility with content integrity.
    \item \textbf{P3: On content (non-obscuring)}: Integrates the label in a way that does not cover any part of the content, ideal for ensuring full visibility of the content while still marking it clearly.
\end{itemize}

\subsubsection*{\textbf{Detail (D):}} This is the provenance of the content in more detail if the content is fully generated by AI, or particularly edited, and when and how it was generated in specifying tools used with dates. The level of detail in a label determines how well users understand the implications of AI-generated content. Detailed labels provide context and action steps, enhancing user education and empowerment. This dimension captures regulatory expectations that call for clear and informative disclosures, ensuring that labels are not overly cumbersome or dismissively brief.

\begin{itemize}
    \item \textbf{D1: Simple warning text}: Provides essential information in a concise format, easy for users to quickly scan and understand, suitable for platforms prioritizing speed and simplicity.
    \item \textbf{D2: Detailed information}: Offers an in-depth explanation about AI involvement, better for educational purposes or platforms where users benefit from more comprehensive data.
\end{itemize}

Covering all 4 dimensions by combinations of possible design options, we derived 10 design samples for warning labels that represent variants of AI-generated content warning labels used by different online platforms (Figure \ref{fig:sample designs}). These design samples were embedded into images where users can see the design within the image as shown in Figure \ref{fig:TRT}.

\begin{figure*}
    \centering
    \includegraphics[width=1\linewidth]{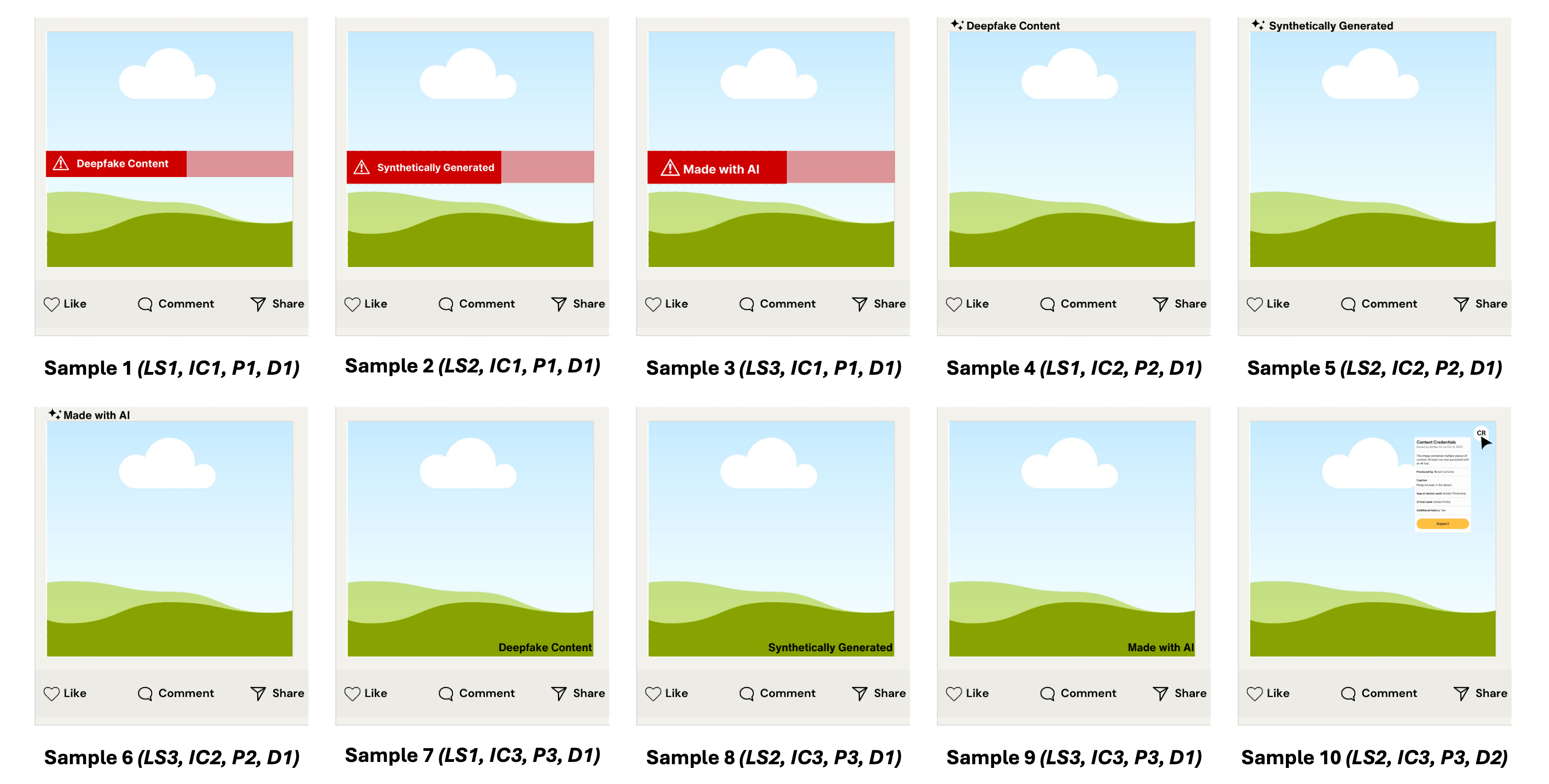}
    \caption{Design Samples derived from design space, showing design options chosen for each in parentheses.}
    \label{fig:sample designs}
    \Description{}
\end{figure*}

\begin{figure*}[ht]
    \centering
    \includegraphics[width=1.00\linewidth]{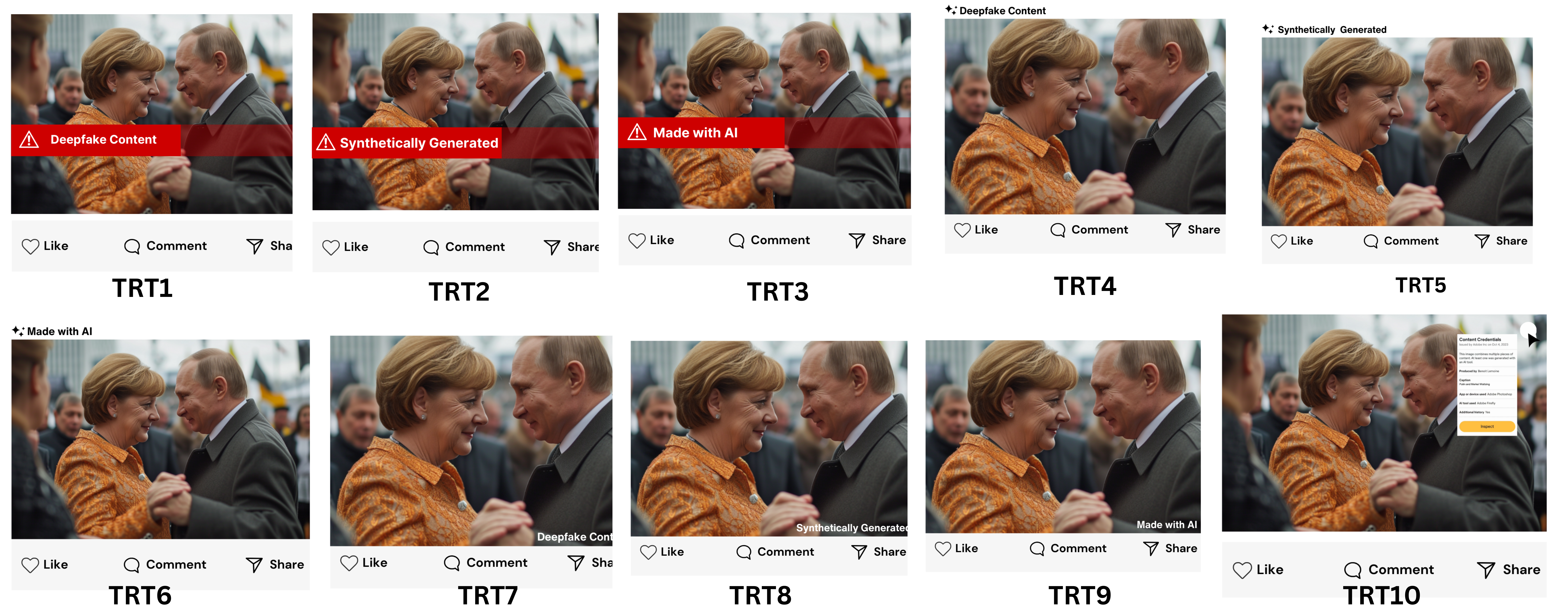}
    \caption{Mean scores for (a) likelihood to engage; and (b) support for making informed decisions to engage}
    \label{fig:TRT}
    \Description{This shows how design samples embedded into a images used to evaluated; in particular this is the Political type of image categorize as a deepfake }
\end{figure*}

Next, we explain the analysis using the measurements collected by the survey in the experiment design using these warning label samples. The designs were embodied to AI set of images that are political and entertainment category. First, we present the results of how users perceive the believability of the content when they see a label--- this is simply the users' trust in content when labeled. Then we present the results of their trust levels for the label designs. Finally, we present the impact of the label on engagements such as like, comment, and share.
\subsection{RQ2: How do Warning Labels Affect Users' Belief of Whether the Social Media Content is AI-generated?}

\subsubsection{Belief that the Content is AI-generated}
We measured \textsc{Belief-Of-Content} using the survey question \textit{"Q3: If you see this on social media, are you likely to believe that this is an image that was created or significantly altered by Artificial Intelligence (AI)? [Likert scale 1 to 5]"}. This question was posed to each participant for the 8 images in each condition (control and 10 treatment groups). To assess the impact of labeling content, we performed an ANOVA between control and treatment groups, to test if there were significant differences between the control condition (no labels) and the treatment groups (TRT1 to TRT10). Subsequently, we performed a within-group ANOVA to check which label gave the participants the highest belief that the content was AI-generated.

Measurements of \textsc{Belief-Of-Content}, in all treatment groups that had a labeled content (TRT1 to TRT10) show significant differences from the control group that did not have a warning label design. The ANOVA results revealed a significant main effect of the treatment group on the believability of the content due to the presence of warning labels \( F(9, 6468) = 4.05 \), \( p < .001 \), \( \eta^2 = 0.006 \). However, the effect size indicates that the treatment group accounts for a very small portion of the variance in trust scores. However, based on the mean values of the measurements, it is evident that having such a label made users believe the content is AI-generated or deepfake.\\
\\
    \fbox{
    \begin{minipage}{\dimexpr\linewidth-2\fboxsep-2\fboxrule\relax}
    \centering
    \textbf{Finding F1:} All label designs led users to believe that the images were AI-generated or edited.
    \end{minipage}}

\begin{figure}[ht]
    \centering
    \includegraphics[width=1\linewidth]{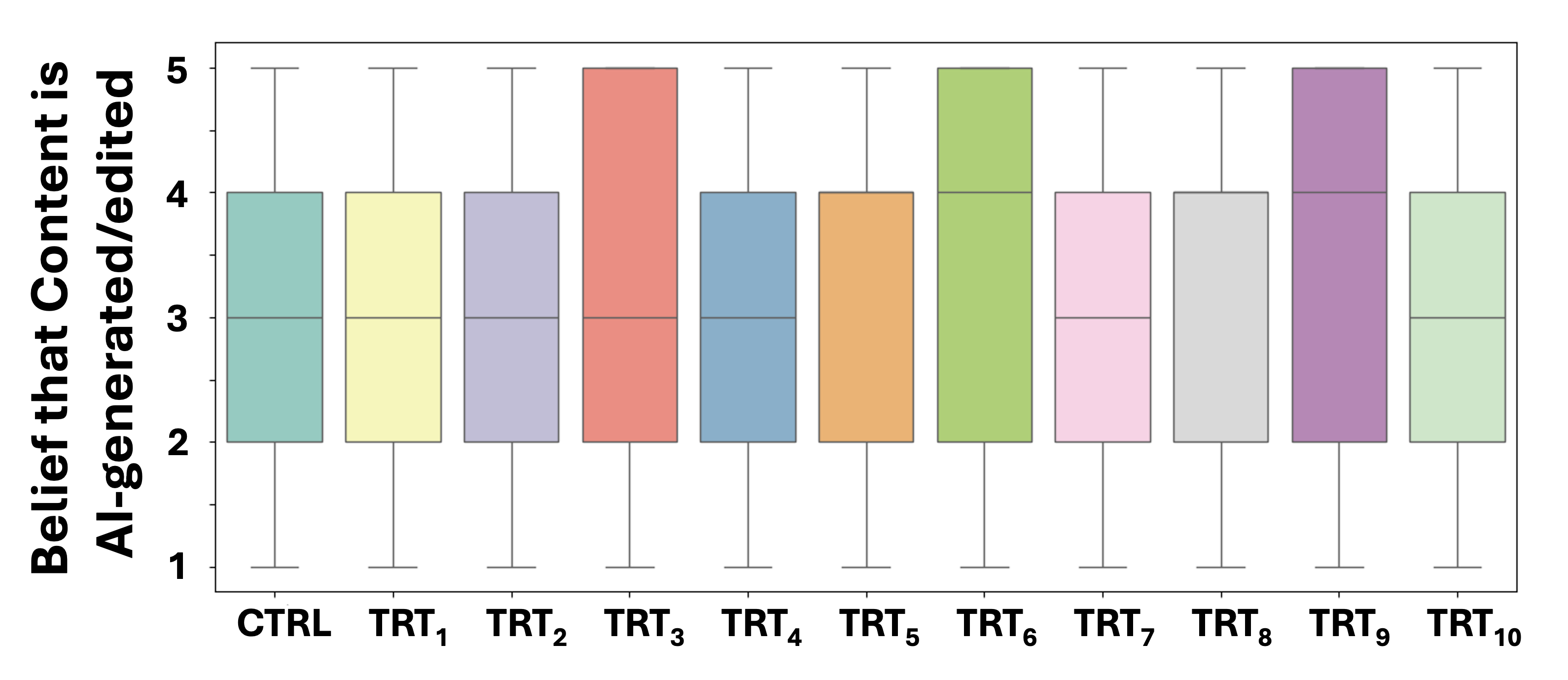}
    \caption{\textsc{Belief-Of-Content} control vs. treatment} 
    \label{fig:Trust to Content}
    \Description{ }
\end{figure}

\subsubsection{Which Warning Label Design Led to Higher Belief that Content was AI-generated?}
To understand which label design had the most effect on users' belief that the content was AI-generated, we examined our results using a between-group ANOVA across the 10 treatment groups. A significant difference between treatment groups was found, \( F(9, 6468) = 4.047 \), \( p < 0.0001 \), eta-squared (\( \eta^2 \)) for the group effect is \( 0.0056 \). A post-hoc Tukey's HSD test was performed to compare the trust scores between the treatment groups. Significant differences between the following groups between TRT1 vs TRT6, \( p = .023 \), TRT1 vs TRT9, \( p = .002 \), TRT2 vs TRT6, \( p = .015 \), TRT2 vs TRT9, \( p = .002 \), TRT4 vs TRT9, \( p = .017 \), TRT7 vs TRT9, \( p = .008 \).  Combining this with an examination of the mean values and effect sizes, we can determine which sample labels led to higher belief in the content being AI-generated.  We found that TRT6 had an effect size of 0.418 and TRT9 had an effect size of 0.445 which is higher than the rest of the treatment groups (see Appendix \ref{sec:beliveability effect size}). Design Sample 9 had the label with the words \textit{``Made with AI''} at the bottom of the image and had the highest belief score (\( F(1, 1332) = 65.97 \), \( p < 0.001^* \), \( M = 3.5 \)). Design Sample 6, using a combination of the words \textit{``Made with AI''} with the diamond icons on top of the image also had a strong effect (\( F(1, 1294) = 57.08 \), \( p = 0.001^* \), \( M = 3.46 \)). Together, these two had the largest difference from the control group, indicating a very strong effect of the label on the perception that the image was AI-generated. Figure \ref{fig:Trust to Content} shows the variance in the believability score against the control group and the treatment groups.

\subsubsection{Does Familiarity Impact the Belief that Content was AI-generated? }
A two-way ANOVA was performed to examine the effect of the treatment group and the \textsc{User-Familiarity} (\(Q_1\)) on the \textsc{Belief-Of-Content} score (\(Q_3\)), as well as the interaction between the treatment group and familiarity score. This revealed a statistically significant main effect of the treatment group on believability scores, \(F(9, 5187) = 5.05\), \(p < .001\), indicating that the treatment group had a significant impact on believability scores. However, the main effect of the familiarity score on believability was not statistically significant \(F(1, 5187) = 0.16\), \(p = .691\), suggesting that familiarity did not independently influence believability scores. Furthermore, the interaction between the treatment group and the familiarity score was not statistically significant \(F(9, 5187) = 1.18\), \(p = .304\), indicating that the relationship between the treatment group and \textsc{User-Familiarity} did not vary based on familiarity (Table \ref{tab:familiarity_interaction}).\\
\\
    \fbox{
    \begin{minipage}{\dimexpr\linewidth-2\fboxsep-2\fboxrule\relax}
    \centering
    \textbf{Finding F2:} Users' familiarity with the content does not affect belief of the content being AI-generated or edited.
    \end{minipage}}
\\

\begin{table*} 
\centering
\caption{Interaction between belief in content being AI-generated and the familiarity with the content}
\label{tab:familiarity_interaction}
\begin{tabular}{ccccc}
\hline
\textbf{Source} & \textbf{Sum of Squares} & \textbf{df} & \textbf{F} & \textbf{p} \\ \hline
C(Treatment\_Group) & 76.61586 & 9 & 5.052351 & 8.11E-07 \\ \hline
\textsc{User-Familiarity} & 0.265351 & 1 & 0.157484 & 0.6915 \\ \hline
C(Treatment\_Group):\textsc{User-Familiarity}  & 17.87401 & 9 & 1.178683 & 0.303798 \\ \hline
Residual & 8739.748 & 5187 & &   \\ \hline
\end{tabular}
\end{table*}


\subsection{RQ3: How Does the Design of the Warning Label Affect Users’ Trust in the Warning?}

\subsubsection{Do Users Trust the Warning Labels?}
We measured \textsc{Trust-In-Label} from our treatment groups asking \textit{ 'TRTQ8-How high is your trust in the label?' [Likert scales 1 to 5]}. A one-way ANOVA was performed within each group to explore the effect of different treatments on trust levels (measured by Q8). The independent variable consisted of ten treatment groups (TRT1 to TRT10) and the dependent variable was \textsc{Trust-In-Label}.   

As highlighted in Table \ref{tab:Trustscore_TRT}, design samples in treatment groups 1, 2, 3, 4, 6, and 9 had significant variances in their trust in the labels. This means that these warning label designs were not equally trusted by all. 
However, warning designs in TRT5 \((F(1,83) = 1.33, p = .232, \eta^2 = 0.014)\), TRT7 \((F(1,80) = 1.19, p = .304, \eta^2 = 0.013)\), TRT8 \((F(1,85) = 0.69, p = .682, \eta^2 = 0.007)\), and TRT10 \((F(1,61) = 0.95, p = .468, \eta^2 = 0.013)\) did not show a statistically significant difference in variance between each other on trust scores to the label. This indicates that although trust varies based on the label design, in design samples 5, 7, 8, and 10 the trust of users did not change significantly.  As depicted in the Figure \ref{fig:Distribution of Mean Trust Scores for Labela and Platform} (a), the highest mean (\(M = 3.44\)) indicated the highest average trust represented in the design sample 10 which had the combination of neutral sentiment without AI and displayed in detail on top of the image. Subsequently, sample 5 (\(M = 3.35\)), sample 8 (\(M = 3.14\)), and sample 7 (\(M = 3.13\)) also showed promising trust levels. To confirm the results in trust with the design samples, we performed a one-way ANOVA between the groups, it showed that TRT10 with warning label design 10 had the highest. We have included all pairwise comparisons and the effect size in the Supplementary file.\\
\\
\fbox{
    \begin{minipage}{\dimexpr\linewidth-2\fboxsep-2\fboxrule\relax}
    \centering
    \textbf{Finding F3:} Design Sample 10, which used the Content Credentials description, elicited the highest trust level.
    \end{minipage}}

\begin{table*}
\centering
\caption{Trust in the label design variance within each treatment group}
\label{tab:Trustscore_TRT}
\begin{tabular}{cccccc}
\hline
\textbf{TRT Group} & \textbf{F-statistic} & \textbf{p-value} & \textbf{Eta Squared} & \textbf{Sample Size} & \textbf{Mean Value} \\ \hline
TRT1  & 3.566323 & \textbf{0.000889}  & 0.035015 & 87  & 3.201149 \\ \hline
TRT2  & 5.733734 & \textbf{1.86E-06}  & 0.057001 & 84  & 3.241071 \\ \hline
TRT3  & 3.997419 & \textbf{0.00027}  & 0.040437 & 84  & 3.136905 \\ \hline
TRT4  & 2.033777 & \textbf{0.048807}  & 0.021498 & 82  & 3.003049 \\ \hline
TRT5  & 1.332669 & 0.231925  & 0.013855 & 84  & \textbf{3.354167} \\ \hline
TRT6  & 4.438919 & \textbf{7.79E-05}  & 0.045757 & 82  & 3.146341 \\ \hline
TRT7  & 1.193841 & 0.303969  & 0.012889 & 81  & \textbf{3.132716} \\ \hline
TRT8  & 0.688545 & 0.681864  & 0.007038 & 86  & \textbf{3.145349} \\ \hline
TRT9  & 5.254580 & \textbf{2.64E-05}  & 0.044372 & 98  & 3.221574 \\ \hline
TRT10 & 0.948736 & 0.468385  & 0.013426 & 62  & \textbf{3.443548 }\\ \hline
\end{tabular}
\end{table*}

\begin{figure*}[ht]
    \centering
    \includegraphics[width=1.0\linewidth]{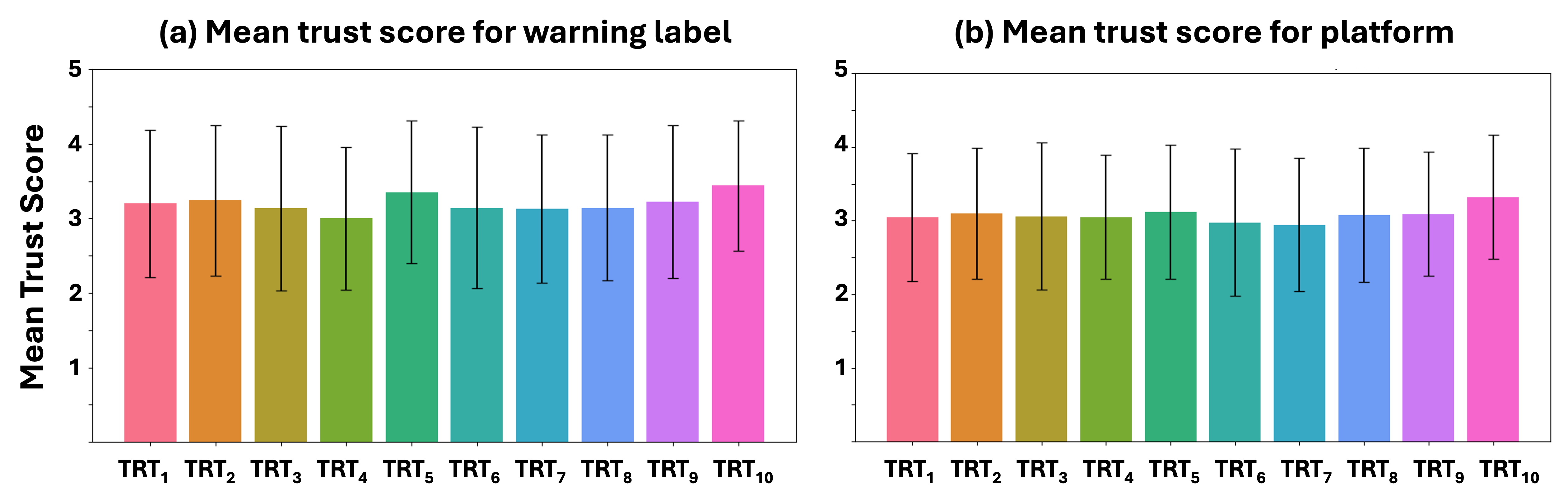}
   \caption{User trust levels for (a) the warning labels; and (b) for the Social Media platform}
   \label{fig:Distribution of Mean Trust Scores for Labela and Platform}
   \Description{ }
\end{figure*}

\subsection{RQ4: Is there a relationship between trust in the warning label and trust in the associated social media platform?}

\subsubsection{What is the User's Trust in the Platform?} We measured \textsc{Trust-In-Platform} from our treatment groups in \textit{ 'TRTQ9-If you see this label on a social media platform, how high is your trust in this platform?' [Likert scales 1 to 5]}. Then a similar analysis was performed for TRTQ9 which evaluates whether the trust in the label translates into the trust of the platform.} The results indicated that none of the treatments showed statistically significant variance within each group(\(p > .05\)). This means all users had a similar attitude towards trust in the platform. Similarly to the previous analysis, design sample 10 had the highest mean (\(M = 3.32\)) (see Figure. \ref{fig:Distribution of Mean Trust Scores for Labela and Platform} (b)).

\subsubsection{Does Trusting the Warning Label Indicate Trust in the Social Media Platform?}
To find if the trust in the label represents the trust in the social media platforms, a Pearson correlation was performed to examine the relationship between \textsc{Trust-In-Label} and \textsc{Trust-In-Platform} between the TRT groups. The scatter plot shows the relationship between Trust in Label (Q8) and Trust in Platform (Q9) across the TRT groups (See fig. \ref{fig:relationshipBtwnLabelandPlatform}). The results indicated a strong positive correlation between trust in the label and trust in the platform, \(r(85)=.73, p < .001\). This suggests that individuals who have high trust in the label also tend to exhibit higher trust in the platform.\\
\\
    \fbox{
    \begin{minipage}{\dimexpr\linewidth-2\fboxsep-2\fboxrule\relax}
    \centering
    \textbf{Finding F4:} The trust in the design of the warning label correlates with trust in the platform.
    \end{minipage}}
\\


\subsection{RQ5: Do warning labels on AI-generated content affect users' engagement behaviors?}

Next, we tried to understand how users are likely to engage in social media posts when there is a deepfake warning label vs without a label. At the same time, we explored whether the content type being political or entertainment affects engagement behavior. 

\subsubsection{Does Having a Warning Label Impact "like" Behavior on Social Media Platforms?}

A one-way ANOVA was conducted to compare the effect of labeling on the likelihood of users to react (e.g. "Like") (\(Q_4\)) to images between the control group (no label) and treatment group (with a label). The analysis revealed no significant difference between the control group (\(n=648\)) and the treatment group (\(n=6,593\)) in terms of likelihood scores, \(F(1,7239)=0.070\), \(p=0.79\). 

\subsubsection{Does Having a Warning Label Impact "Comment" Behavior on Social Media Platforms?}

The same analysis was conducted to compare the effect on the likelihood of users making a "Comment" (\(Q_5\)) on the images between the control group (without a label) and the treatment group (with a label). Similarly, there were no significant differences between the control group (\(n=648\)) and the treatment group (\(n=6,581\)) in terms of \(Q_5\) scores, \(F(1,7227)=0.092\), \(p=0.762\). 

\subsubsection{Does Having a Warning Label Impact "Share" Behavior on Social Media Platforms?}
A one-way ANOVA on the likelihood of users sharing images (\(Q_6\)) revealed no significant differences between the control group (\(n=648\)) and the treatment group (\(n=6,611\)) in terms of likelihood scores \(F(1,7257)=1.404\), \(p=0.236\). This indicates that the behavior of users reacting with 'likes', commenting, or sharing was not significantly influenced by the presence or absence of a label.\\
\\
    \fbox{
    \begin{minipage}{\dimexpr\linewidth-2\fboxsep-2\fboxrule\relax}
    \centering
    \textbf{Finding F5:} Presence of labels did not affect engagement with content (e.g., ``Like'', comment, and share)
    \end{minipage}}
\\

\subsubsection{Does User Engagement Vary Depending on Whether the Content is Political or Entertainment? }
\subsubsection*{\textbf{Like behavior}}We performed a two-way ANOVA to examine the interaction between content category (\textsc{Political} vs. \textsc{Entertainment}) and labeling (\textsc{Control} vs. \textsc{Treatment}) on participants' likelihood to react ("Like") to images. There was a statistically significant effect of category on \(Q_4\) scores, \(F(3,7237)=3.81\), \(p=0.01\), suggesting that different categories (political vs. entertainment) and treatment conditions (control vs. labeled) significantly influenced participants' reactions (See Figure. \ref{fig:comparisonQ4Q5Q6} (A), and Table. \ref{tab:AnovaQ4}). A Tukey's HSD post-hoc test (listed in Appendix \ref{sec:Posthoc Test in Q4 Likelihood to like}) suggests that the main difference driving the overall significant result in the ANOVA is the difference between the \textsc{Entertainment}\(_{Treatment}\) and \textsc{Political}\(_{Treatment}\) groups. 

\subsubsection*{\textbf{Comment behavior}} The two-way ANOVA on participants' likelihood to comment on images (\(Q_5\)) showed a statistically significant effect of content category on likelihood to comment scores, \(F(3,7225)=16.26\), \(p<0.001\). The post-hoc Tukey's HSD test results showed significant differences between the same \textsc{Entertainment}\(_{Treatment}\) and \textsc{Political}\(_{Treatment}\) groups (see \ref{sec:Posthoc Test in Q5 Likelihood to Comment}, suggesting that when labels are applied political images were significantly more likely to be commented on compared to entertainment images (See Figure. \ref{fig:comparisonQ4Q5Q6} (b) and Table.\ref{tab:AnovaQ5}).

\begin{table*}[ht]
\centering
\begin{tabular}{lcccccc}
\toprule
Category                & Mean Q4 Scores & Std Q4 Scores & Sample Size & DF Between & DF Within \\
\midrule
Entertainment Control    & 1.632716        & 0.997354      & 324         & 1          & 323       \\
Entertainment Treatment  & 1.614545        & 0.994940      & 3300        & 1          & 3299      \\
Political Control        & 1.700617        & 0.985907      & 324         & 1          & 323       \\
Political Treatment      & 1.696629        & 1.052384      & 3293        & 1          & 3292      \\
\bottomrule
\end{tabular}
\caption{ANOVA Summary for Political and Entertainment Groups on for Q4 Likelihood "Like" react}
\label{tab:AnovaQ4}
\end{table*}

\begin{table*}[ht]
\centering
\begin{tabular}{lcccccc}
\toprule
Category                & Mean Q5 Scores & Std Q5 Scores & Sample Size & DF Between & DF Within \\
\midrule
Entertainment Control    & 1.425926        & 0.769428      & 324         & 1          & 323       \\
Entertainment Treatment  & 1.389701        & 0.744086      & 3282        & 1          & 3281      \\
Political Control        & 1.475309        & 0.792383      & 324         & 1          & 323       \\
Political Treatment      & 1.531979        & 0.919662      & 3299        & 1          & 3298      \\
\bottomrule
\end{tabular}
\caption{ANOVA Summary for Political and Entertainment Groups on likelihood to comment (Q5)}
\label{tab:AnovaQ5}
\end{table*}

\begin{table*}[ht]
\centering
\begin{tabular}{lcccccc}
\toprule
Category                & Mean Q6 Scores & Std Q6 Scores & Sample Size & DF Between & DF Within \\
\midrule
Entertainment Control    & 1.845679        & 1.335725      & 324         & 1          & 323       \\
Entertainment Treatment  & 2.042470        & 1.501733      & 3320        & 1          & 3319      \\
Political Control        & 2.225309        & 1.542537      & 324         & 1          & 323       \\
Political Treatment      & 2.177150        & 1.542875      & 3291        & 1          & 3290      \\
\bottomrule
\end{tabular}
\caption{ANOVA Summary for Political and Entertainment Groups likelihood of sharing (Q6) }
\label{tab:AnovaQ6}
\end{table*}

\subsubsection*{\textbf{Sharing behavior}}
A two-way ANOVA was performed to examine the interaction between the category of content (\textsc{Political} vs. \textsc{Entertainment}) and the labeling (\textsc{Control} vs. \textsc{Treatment}) on the likelihood of participants sharing images (\(Q_6\)). There was a statistically significant effect of category on (\(Q_6\)) scores, \(F(3,7255)=8.21\), \(p<0.001\). This indicates that the likelihood of the participants sharing images varied significantly depending on the content type and whether the images were labeled or not. However, when we tried to understand which groups had the difference by using Tukey's HSD post-hoc test (Listed in Appendix \ref{sec:Posthoc Test in Q6 Likelihood to share} B.3), we found that a significant difference was found between \textsc{Entertainment}\(_{Control}\) vs. \textsc{Political}\(_{Control}\), \textsc{Entertainment}\(_{Control}\) vs. \textsc{Political}\(_{Treatment}\), and \textsc{Entertainment}\(_{Treatment}\) vs. \textsc{Political}\(_{Treatment}\). However, no differences were found between \textsc{Entertainment}\(_{Control}\) vs. \textsc{Entertainment}\(_{Treatment}\), \textsc{Political}\(_{Control}\) vs. \textsc{Political}\(_{Treatment}\), and \textsc{Entertainment}\(_{Treatment}\) vs. \textsc{Political}\(_{Control}\). 

\begin{figure*}[ht]
    \centering
    \includegraphics[width=1.00\linewidth]{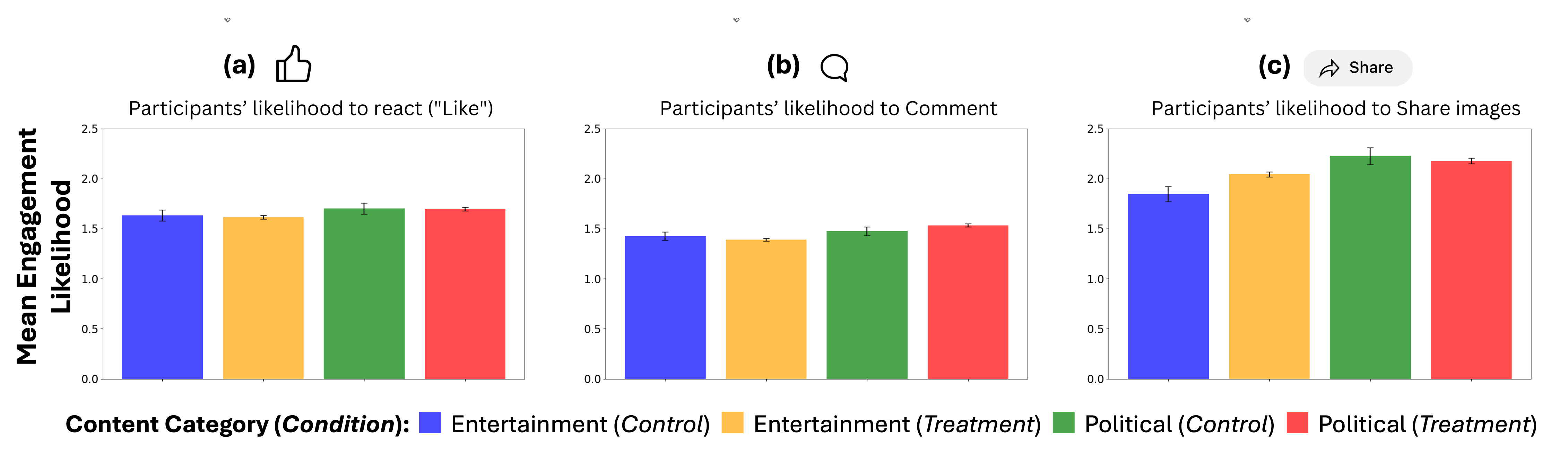}
    \caption{Mean engagement likelihood by content category for (a) ``Like'' reaction; (b) commenting; and (c) sharing}
    \label{fig:comparisonQ4Q5Q6}
    \Description{(A). Mean Responses for Q4 Likelihood "Like" react (Engagement Score) where the only significance
found in Entertainment(Treatment) and Political(Treatment)); (B). Mean Responses for Q5 Commenting Likelihood
(Engagement Score) found only to be significant in Entertainment (Treatment) and Political(Treatment)  (C). Comparison of Sharing Likelihood Based on Label and Content Type (Q6) found a significant difference between Entertainment(Control) vs. Political(Control), Entertainment(Control) vs. Political(Treatment), and Entertainment(Treatment)  vs. Political(Treatment)}
\end{figure*}

Importantly, this suggests that users' sharing behavior of political vs. entertainment content remains the same whether there is a label or not. Although having labels appeared to reduce sharing behavior in political images and increase sharing in entertainment images (as shown in Figure. \ref{fig:comparisonQ4Q5Q6} (c), and Table.\ref{tab:AnovaQ6}), these differences were not statistically significant, as confirmed by the post-hoc test results. In other words, any label design may not have a significant impact on users' desire to share images. However, according to our label designs, the highest likelihood of sharing was found in Sample Label Design 10, which had the Content Credential Descriptions on the images, and the lowest was in Design Sample 6, which had the wording "Made with AI" on the image.\\
\\
    \fbox{
    \begin{minipage}{\dimexpr\linewidth-2\fboxsep-2\fboxrule\relax}
    \centering
    \textbf{Finding F6:} The content category changed the sharing behaviors of users, whether the content had a label or not.
    \end{minipage}}
\\

\subsubsection{Did the warning labels help users think before they engage with the content?}
A one-way analysis of variance (ANOVA) was performed to compare the effect of different treatment groups (TRT1 to TRT10) on engagement scores based on whether any label in the image made them think before engaging. This was measured using \textit{ 'TRTQ10-Seeing this label in the image made me think before I engage (react-like/heart/wow, comment , share) with it?' [Likert scales 1 to 5]}. Although the ANOVA showed significant differences in engagement scores between at least some of the treatment groups \( F(9,6499) = 1.89 \), \( p = .048 \), Tukey's HSD post-hoc test revealed that the only difference was observed between TRT2 and TRT6. The overall mean responses in the treatment groups range from approximately 3.34 to 3.55 on a 1-5 likert scale (See \ref{sec:Q10 seeing}). This suggests that on average, participants tended to somewhat agree with the statement in all treatment groups. The mean responses are close to the midpoint of the scale, and the small effect size \( \eta^2 = 0.00261 \) indicate that the differences between the groups don't suggest a major impact on user behavior.

\subsubsection{Did the warning labels help users make informed decisions before they engage with content?}
\begin{figure*}[ht]
    \centering
    \includegraphics[width=1.00\linewidth]{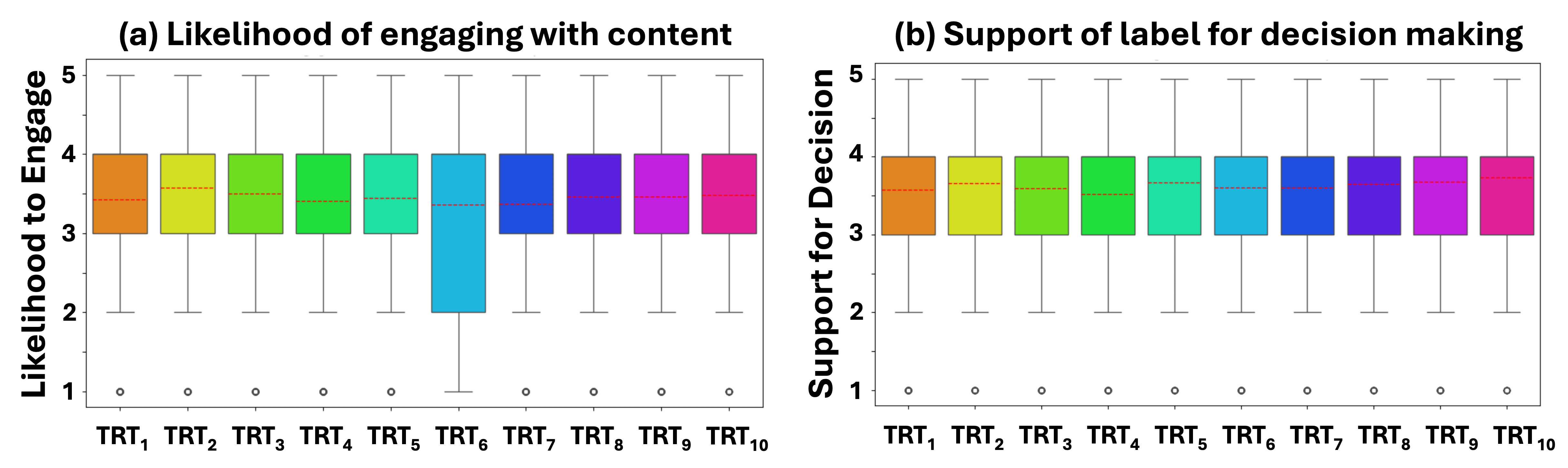}
    \caption{Mean scores for (a) likelihood to engage; and (b) support for making informed decisions to engage}
    \label{fig:Q10Q11Mean}
    \Description{(A). Mean Responses for Q10 (Engagement Score) and (B) Mean Responses for Q11 (Decision-Making Score)}
\end{figure*}

Similarly, on the question \textit{ TRT11-'This label helped me to be informed about the image where I can make informed decisions on how I engage with it.  [Likert scales 1 to 5]} ANOVA results revealed a statistically significant difference between the treatment groups \( F(9, 6593) = 2.04 \), \( p = .031 \).  The post-hoc test suggested that the only statistically significant difference was between TRT2 and TRT4 \( (p < 0.05) \). The effect size for the ANOVA results, calculated using Eta Squared, is \( \eta^2 = 0.00278 \), see Figure.\ref{fig:Q10Q11Mean}, (a) and (b). 

There is only approximately 0.26\% - 0.28\% of the variance in engagement scores (\(Q_10\), \(Q_11\)) which can be explained by the differences between the treatment groups (TRT1 to TRT10). This implies that although the differences are statistically significant, the practical impact or real-world effect of treatments on decision-making is minimal. In other words, it is evident that although all users somewhat agreed that labels helped them think before they engage and make informed decisions, labels in the image had a generally mild impact.\\
\\
\fbox{
    \begin{minipage}{\dimexpr\linewidth-2\fboxsep-2\fboxrule\relax}
    \centering
    \textbf{Finding F7:} All label designs helped users to think and to make informed decisions before they engaged with content but overall, the impact is very mild.
    \end{minipage}}

\subsection{Qualitative Findings of the Survey}
Our experiment collected responses to some open questions where users could share their opinion freely.

\subsubsection{What are the AI content encountered by users and where have users seen them?}
\begin{figure}
    \centering
    \includegraphics[width=1.0\linewidth]{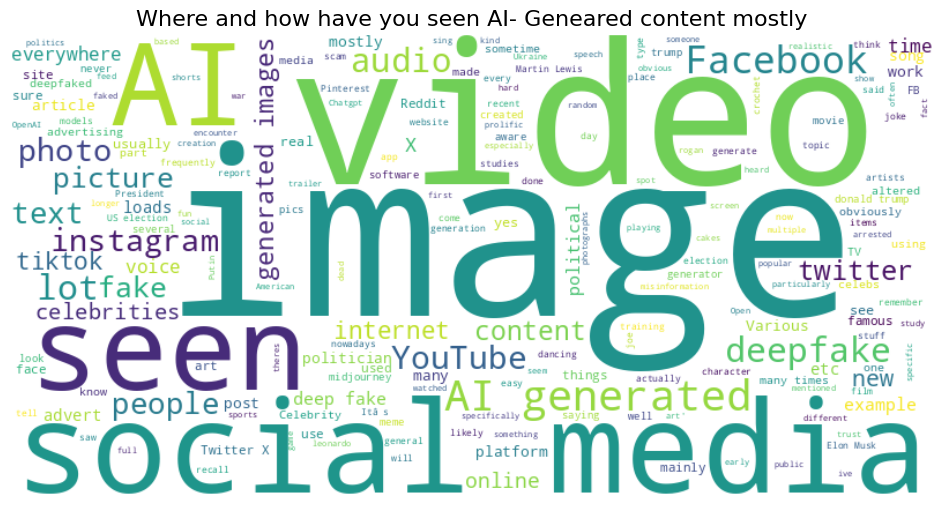}
    \caption{Ward-cloud capturing how Users encounter deepfake or AI generated content---they found mostly in the form of Videos, images and most of them seen in Social Media}
    \label{fig:wordcloud}
\end{figure}

We asked all participants whether they have encountered deepfake or AI-generated content before (e.g., text, image, video, audio) and explain how they have seen it in a open text field. Our results show participants listing many social media platforms with particular events of seeing deepfakes---i, e ``\textit{Molly Mae's voice being faked on TikTok following her break up with Tommy'', ``I've encoutered AI generated boyfriend/girlfriend, after my eldest daughter interacted with it'', ``Pope in a puffa jacket image'' }. We had 845 meaningful responses in the text entries. Based on the frequency of words which they encountered we made a word cloud to identify key elements of how users confront AI-generated content. The Fig. \ref{fig:wordcloud} indicates that participants found AI content in videos, images and mostly in Social media.

\subsubsection{What are these label designs meant and how to improve them? }
In another entry, we asked users to explain how they feel about the designed warning labels and to provide suggestions on how they think the labels could be improved. Answers to this text field provided us the opportunity to explore why some design samples were trustworthy and some were not.

The labels communicated to the users that the images are indeed either deepfake or AI generated\/edited. The presence of labels on real images made users believe that even these were AI generated. Interestingly, the Sample 9 (had wording 'made with AI' positioned to the bottom right)  and Sample 6  (had wording "made with AI and 'diamond' icon which is used by some online platforms to indicate the use of generative AI technologies) outperformed in communicating the message with positive comments. Some reviews indicate why it is best design as \textit{ 'the label was clear and obvious, I don't think it needs to be improved'} and some suggestions such as \textit{“Colour could be bolder”} ,  \textit{“it is helpful , but it would be good to know how much AI has done, for example, with links provided to the originals on which it is based'}. At the same time, some users had criticized the potential danger in which people can manipulate these labels by stating \textit{ 'Useful, maybe a watermark also?' So it can't be cropped out”}. 

It was evident that the effect of including a label was to influence users to think that the content is AI generated. Some of our designs were not found to be clear when we used the term ``Deepfake''. It was clear from the comments of the participants on Design Sample 6 (Made with AI with diamond icon on top) - \textit{ ``I would assume that the label is correct and telling the truth. Although I have seen these images before and I know not all of them are fake''}, \textit{``It made me consider that the photos I thought maybe weren't AI actually were AI''}. Design sample 1 which had red icons and words deepfake had comments as \textit{``Label said Deepfake but did not mention AI.'', ``I think 'deepfake' doesn't work as well as the term 'AI Generated' but it gets the job done at least. I haven't seen this kind of label on the web yet but they should exist, but I think the only way for them to work would be crowdsourcing since people who upload AI Generated content often claim that it is real.''The label didn't say it was AI generated it said it was deep fake.   The label should SUGGEST that it might be AI generated''.}

Most of our label designs affected the trust scores within the group. Design Sample 10, which had the Content Credentials icon with description of the AI generated content similar to nutrition information labels in food packaging, had highest trust score as a label. We examined comments from users on explaining their choice to trust. Some participants explained \textit{“Specifying which part of the photo has been altered.”} , \textit{“Clear and concise”}, \textit{“smaller and then clicked on to get more info”}. However, some were very concerned about the information overload that could potentially ignore the main purpose of the label. Comments made by participants said \textit{``CR label - the information should be better presented rather than in the many points, as there was too much info so it was easy to ignore'', ``As the label say it combines multiple pieces of content, it doesn't make clear what part is AI. It is very hard to improve the lablels.'', ``You look at the image first each time and the label second. There's a lot to read on the label so most people might skip it and not be as informed as the label was trying to encourage.''} 
We had nearly 900 free text responses across all participants on each label in each image. In general, this feedback highlights the need for a clear, informative, and trustworthy labeling of deepfake content to help users make informed decisions and mitigate the spread of misinformation. Our team collectively made open codes to categorize the issues and opportunities of each designed sample.  We categorize these codes into five themes and present these findings below.

\subsubsection{Clarity \& specificity:}
 Participants expressed a desire for labels to explicitly state "AI-generated" or "AI-manipulated" rather than "deepfake," and to specify the alterations made to the image. They found the term deepfake to be ambiguous. It was clear when participants highlighted as \textit{``the label should have written on it AI-generated'', ``The images were labeled as 'Deep Fake' which makes me think that they could have been created by a human or AI.'', "Deepfake is almost an outdated term, needs to be clearer whether it was AI-generated altered (Photoshop) or something similar"}. On the other hand, some participants made recommendations such as \textit{``I think the language attached to the label needs to be extremely clear - using the word 'fake' where possible, or 'heavily edited' - as that is terminology that people understand more than 'credentials'”}.
 
\subsubsection{Visibility \& placement:}
Participants explained that some labels attracted more attention by being highly visible yet were ambiguous about their intent. Many participants suggested reducing visibility, making them less intrusive, or placing them in a different location on the content. For instance \textit{`` "Label needs to be smaller and lower'', ``it should not be that big and cover a big part photo. Should be labeled either in a corner of the photo in the posted bio, or outside the image.'', ``The labels could be smaller and in one corner just to be able to see the picture properly. They should still be visible though indicate that''}. This shows that participants preferred visible labels that did not obstruct the content. 

\subsubsection{Additional information:}
 Participants proposed including details about the image's source, the AI software used, or an explanation of how the image was altered. The participants expressed the need to quickly provide the information in one glance without having to read more details. Some designs led to more cognitive load to understand if the image is generated or edited. Some icons, such as CR (content credentials) did not immediately provide enough information at a glance, needing another click which seems a pain point. They suggested being clear upfront and providing details additionally when clicking. \textit{``CR wasn't clear what it meant but we needed to see more details to judge''. ``The labels are informative.  However, it could be made clear if the image was "created" or "altered" by AI.''} 

\subsubsection{Trustworthiness \& authenticity:}
Concerns were raised about the legitimacy of the labels themselves and the need for a reliable verification process. Specifically, Design Sample 1, which had the warning <!> icon in red, obstructing the content, with the wording ``Deepfake content'' had the lowest trust score. Some participants wondered if the label itself was faked. Participants expressed concern that some designs can be easily generated and edited by anyone to manipulate. \textit{`` someone can easily create this label, and even add to real images to make it AI'', ``Less intrusive it looks like a fake label so I would trust the image more than the label'', ``deep fake is almost an outdated term, needs to be clearer whether it was ai generated altered (Photoshop) or something similar'',``Label said that was a Deepfake, not AI-generated. I think images should be grayed out and then by clicking a disclaimer you can view the AI/Deepfake image.''}

\subsubsection{Alternative methods for labels:}
Suggestions included using watermarks (\textit{``Perhaps a watermark type of label would make it harder to remove.''}), blurring the image until the warning label is clicked (\textit{``Maybe the image is blurred if it's 'Deepfake content', and you have to click on it to see it.''}), or incorporating a logo to indicate AI-generated content (\textit{``would prefer some form of logo''}).

\section{Discussion}

Labeling AI-generated content is essential to combat the risks of misinformation and ensure transparency. Since AI technologies can produce highly realistic content, such as deepfakes or fabricated content that can deceive users, society faces significant challenges in being able to trust online content \cite{xu2023combating}. Without clear labels, users can mistake AI-generated content for human-created material, leading to misinformation or confusion about the authenticity of the content. This is especially important because AI-generated misinformation can have significant social, political, and economic consequences. Researchers emphasize that labeling can help users critically evaluate the information they encounter, mitigating the risk of misinformation being spread inadvertently \cite{molina2021fake, papakyriakopoulos2022impact}. Beyond misinformation, in the context of AI models dominating large-scale content generation, efficient labeling promotes transparency and accountability on platforms. On the other hand, it is vital for users to understand when they are interacting with AI-generated content to adjust their expectations and judgments accordingly. As we see exponential growth of AI-generated content there is a need to better understand how users perceive and engage with this. However, due to the rapid movement of regulatory acts, some industries are making efforts to increase the transparency and accountability of such content using visible (or invisible) watermarks on the content. In this work, we provided a design space for warning labels for AI-generated content and explored the users' engagement and their trust in labels derived from this space using empirical tests. We had two main objectives from our Design Samples chosen as warning labels to evaluate: 1) to communicate by a label on content that was generated or altered by AI, which will lead to informed decisions about the content, and 2) to build trust, which will reduce misleading effects in user engagements. In this section, we discuss how our design approach helped to achieve these goals and, based on the experiment results, how the design needs to change. Finally, we discuss the wider implications of our work and limitations in this study.

\subsection{Communicating warnings about AI-generated content}
Warning labels for AI-generated content should communicate clear, concise, and actionable information to users. The effectiveness of these labels depends on how well they inform users about the nature of the content, its origin, and any potential associated risks \cite{miranda2019automated}. We tried to capture many communication factors in our label designs. Specifically, our designs combined features from four main dimensions that best describe the label---Label sentiment, icon or colors, positioning of the label, and level of detail provided. Our design samples had a variety of levels and direct language to state whether the content was AI-generated. In some samples, we embedded some warning icons that indicate risk warnings to prompt users to approach content with caution. Some had visual markers (icons) for quick recognition. Some samples contain detailed descriptions, which help users further evaluate the authenticity of the content. We placed our labels in a few positions to understand the efficacy of contextual placement of labels to ensure visibility, as explained in our designed warning label samples (RQ2). {Our results indicate that warning labels with a hazardous tone that had the term ``Deepfake'' were consistently less effective. This is mainly due to the ambiguity it caused and the users much preferred ``Made with AI'' or language that included ``AI'' in it. Users preferred neutral labels with icons commonly seen in an AI context (e.g., \includegraphics[width=0.3cm]{aiicon.png}) or without an icon. In previous research, text was found to be more prominent than icons in security communications ~\cite{stransky2021limited}. In terms of positions, users found that obstructing the image with color is disturbing the content and the preference was for the label to be positioned above or below the image. Design Sample 6, which included the commonly used neutral icon, led users to have a higher belief in the content. Interestingly, in terms of informing the user about content being AI-generated or deepfakes, Design Sample 1, 2, 3 4, 7, and 10 did not perform well. Our interpretation is that this was mainly caused by the ineffective communication to the user about what the label intended. Although we see Design Sample 10 (which used the Content Credentials icon) performed well in building trust on the label, it did not perform well by bringing the message to users on AI-generated content ( see effect size in \ref{sec:beliveability effect size}). Further, we found that users prefer to see the CR icon to provide some words or icons to communicate if it was AI-generated than having to read more details on content details and edits on second-level interactions. Many users are concerned about information overload due to the level of detail indicated in such designs. This resonates with research that has shown that strategies on detail-level provenances backfire due to perceived accuracies in the content~\cite{feng2023examining}.

\subsection {Enhancing Users' Trust }
In a digital landscape increasingly shaped by AI, public trust in the general information ecosystem depends on the integrity of mechanisms such as warning labels. Trustworthy labels help preserve the integrity of news, social media, and other content-sharing platforms. If the public loses trust in these labels, it can lead to widespread skepticism about the authenticity of all content, eroding confidence in the digital space as a whole. In our research, we tried to make labels reliable and trustworthy by adding warning icons, colors, and details for descriptions of the content. From our results, it is clear that, for warning labels to be effective, users need to trust that they are accurate and applied consistently. If people doubt the legitimacy of a label, they may ignore it, undermining the very purpose of informing the public about AI-generated content. A trustworthy label system builds credibility, allowing users to trust the information provided and make informed decisions. Our results showed all label designs with wording in a hazardous tone and icons in red, such as Design Sample 1, 2, and 3 had significant differences in trust of the label (See Table \ref{tab:Trustscore_TRT}). 

The design samples we created using design space showed trust towards the platform as well. Our tests showed no significant variances within their groups on their trust in the platform. The results on correlation proved this with a strong coefficient. This finding highlights that, if users trust the label, it leads them to trust the social media platform that labels the content. Currently, there are many models to label misinformation on social media platforms--- e.g., community labels, fact checks by professional organizations, and automatic labels. With the recent introduction of regulatory requirements for labeling AI content, we are witnessing many labels on social media platforms. However, it is unclear how these labels are perceived by the public. We asked our users to make choices on the question \textit{“Who do you think is most suited to create and tag the labels you saw in this experiment?”}. Our results suggest that many users felt that the labeling should be processed by the tools themselves or the platform itself. At the same time, for systems that host or distribute content such as social media platforms, trust in AI-generated content labels is essential to maintain their accountability. If users trust that platforms are accurately labeling AI-generated content, it reassures them that the platform is taking steps to prevent the spread of misleading or deceptive information. In contrast, mistrust in labels could lead to a loss of confidence in the platform’s commitment to curbing misinformation. Therefore, articulating trust in labels should be done with care. Strategies involved in creating trustworthy warning labels depend on our first object, where it is needed to have a clear communications strategy. Using consistent and visible labeling across platforms, prioritizing transparency, and user engagement, stakeholders can create a more trustworthy labeling system that empowers users to navigate AI-generated content. However, labels alone are not a panacea and should be considered one component of a broader effort to ensure trustworthy AI. Enhancing AI literacy by educating the public about AI technologies and their implications is worth considering. This empowers users to critically evaluate content and builds trust in the labeling system.

\subsection{Implication of the Study} 
Our work demonstrates that AI-generated content labels play a critical role in influencing user perception, trust, and engagement with digital content on social media. Although based on our sample (n= 911), the overall impact of labels on engagement is mild, the design and context of these labels bring important insights into how users perceive and react to such labels. This enables both HCI designers and practitioners to design effective interventions which build trust and shape user behavior~\cite{gao2018label}. The results also inform policymakers, helping improve alignment between emerging legal requirements in regulating AI technologies and the design of warning labels. Currently, the European Union's AI Act, calls for transparency in the deployment of AI systems~\cite{EuAIact}.  Our research had 7 direct findings from which implications can be drawn to design effective user interfaces that show warning labels to enables users to make informed decisions, thereby enhancing the platform's trust and accountability. Moreover, it will address the broader societal challenges posed by deepfakes and AI-generated content as discussed below. 

\subsubsection{Designing for Trust in Labels and Belief in the Content}

Our finding (F1) reveals that all the label designs led users to believe that the images were deepfakes or AI-generated/edited. This suggests that labels play a crucial role in shaping user perception, even if the content is not necessarily AI-generated. The mere presence of a label prompts users to question the authenticity of the content. This has significant implications for the transparency of AI content, as labels must be carefully designed to prevent unnecessary skepticism or misinterpretation. The design options we used in the framework dimensions had varied efficacy on each. Overuse of labels might lead to confusion or warning fatigue, where users start doubting the credibility of all content. This aligns with practices recommended in research that shows the need for transparent communication of AI's role in content creation to avoid misunderstanding~\cite{Mozilla}.

\subsubsection{Effective Labels are Not Impacted by Familiarity with the Content}
According to finding (F2), users’ familiarity with the content did not influence their belief in the label’s assertion that the content was AI-generated. This finding suggests that labeling strategies can be universally applied regardless of the user's prior knowledge or familiarity with the topic. This highlights the robustness of generative AI warning labels as a tool for combating misinformation and disinformation, particularly in cases where users are less informed or are engaging with unfamiliar content. Moreover, it indicates that labels work consistently across different user demographics and content types, a valuable insight for platforms aiming for wide-scale adoption of such strategies~\cite{pennycook2020implied}.

\subsubsection{Careful Design Affects Trust in Labels and Impacts Platform Credibility}
Findings (F3) and (F4) highlight the importance of label design in building user trust. The specific label Design Sample 10 with Content Credentials description garnered the most trust, underscoring that not all label designs are equally effective. This suggests that design elements such as the inclusion of content credentials or metadata play a pivotal role in enhancing credibility. However, it was not the label that was most suitable for communicating the label's intention. This is a significant finding in our research as users prefer the label to be upfront and clear in its message. Furthermore, (F4) demonstrates that the trust placed in the label translates directly into trust in the platform hosting the content. This is crucial for social media companies, news outlets, and other platforms, as their credibility hinges on effective communication with users about content authenticity. Poor or inconsistent labeling can erode trust, which may harm the platform’s reputation.

\subsubsection{User Engagement is Dependent on Content Type}
Findings (F5) and (F6) indicate that labels did not affect overall engagement (e.g., likes, comments, sharing). This is a somewhat different behavior from normal misinformation behaviors when there is a warning label. Previous research found that warning labels reduce sharing or engagements and are effective \cite{martel2023misinformation}, but for AI-generated images, these effects seem different. However, we found that the nature of the content (political or entertainment) led to different user reactions. In the political context, labeled content prompted more reactions and comments, possibly because users are interested in politically sensitive information. In contrast, entertainment content did not attract the same level of scrutiny. This suggests that platform operators should tailor their label strategies depending on content type, recognizing that users are more discerning when interacting with high-stakes content, such as political content. It also emphasizes the need for nuanced approaches when addressing misinformation in different domains.

\subsubsection{Influence to Sharing and Decision Making}
Despite the overall mild impact noted in Finding (F7), all label designs somewhat encouraged users to pause before engaging with content, whether by liking, commenting, or sharing. This aligns with prior research suggesting that users are more likely to make informed decisions when prompted with cues to reflect on the authenticity of the content \cite{pennycook2021shifting}. Although the impact was not overwhelming, even small behavioral changes can have significant implications for the spread of AI-generated disinformation. This suggests that labels may be an essential tool in promoting more responsible social media behavior, even if their effects are subtle.

\subsection{Limitations and Future Work}
Although our study on AI-generated content warning labels provides valuable information, we acknowledge the challenges and limitations in our design and analysis of the study within and between subjects. 
Warning labels can have diverse designs to meet platform and users' needs. In our paper, we set four dimensions and created design samples that may be potentially biased to our team's judgment. However, as research highlights, there are a variety of strategies by which these designs can be created~\cite{feng2023examining}. One is bringing more variations to the detailed design options we identified. In addition to designs that provide provenance and authenticity by C2PA \cite{C2PA} there are many other standards that try to implement authenticity, such as Four Corners \cite{fourcorner} which has cryptographic signed chains and other provenance standards. Our design dimensions provide an initial direction to how effective warning labels can be provided for AI-generated content in terms of transparency and communication, where a greater variety of labeling options can be tested. In future studies, we plan to expand and extend to more designs and evaluate them. 

Our study had small effect sizes that challenged the applicability of warning labels for AI-generated content in real-world settings. The study was conducted in a controlled environment that does not fully capture everyday user behavior, where factors such as warning fatigue, external influences, and varying engagement across different platforms are included. We removed responses from participants who failed our attention check from our analysis even if they completed the test. In addition, participants in such studies may be more attentive or cautious than typical users, which can skew the findings, especially when it comes to assessing trust and engagement with labeled content.  Our repeated statistical analysis may affect the significance test, and also the confounding factors may not be clear in this setting. We plan further analysis using linear mix models instead of ANOVA which will aim to address variable effects. 

While these studies show that label design impacts user trust, the effects are often short-term and may not reflect long-term behaviors or changes in user attitudes. The diversity of user demographics, content types, and external variables in real-world scenarios makes it difficult to generalize the findings of a controlled experiment. The data from this research can be further analyzed to investigate how each group performs in relation to their internet behavior preference. Specifically, the findings on sharing behaviors can be further analyzed to assess participants' accuracy in detecting deepfakes, and if that accuracy has any relationship to their sharing intentions. This could be useful because one of the objectives of the warning label is to mitigate any negative post-effects after seeing deepfake content.

We recommend further research that takes into account these complexities by testing labels in more natural environments with diverse participants to fully understand their effectiveness and impact. We specifically expect to conduct more nuanced studies in designing warning labels for marginalized communities and users in the global south on understanding how to implement transparency and authenticity. This will enable the design of effective warning labels that help mitigate the risks inherent in a world full of AI-generated content.

\subsection{Conclusions}
In conclusion, we have explored the design space for warning labels for AI-generated content and conducted an experiment to test the efficacy of sample label designs which demonstrates label design options need to be carefully chosen. Although the study revealed that labeling influences user perceptions and behaviors, its overall impact on engagement is relatively mild. The presence of labels significantly increases users’ belief that content is AI-generated or altered, regardless of their familiarity with the content, and builds trust in the platform hosting the content. In particular, specific label designs, such as Sample 10 with Content Credentials, are more effective in fostering trust than others, but lack the performance in communicating the message to users about AI-generated images. Labels with a neutral icon such as \includegraphics[width=0.3cm]{aiicon.png} and words with ``AI'' performed well in communicating the message. However, despite these influences, labels alone do not dramatically alter user behaviors, such as sharing or commenting, particularly when the content is political or entertainment-oriented.

The broader implications of this research suggest that while labels are a valuable tool in raising awareness about AI-generated content and encouraging informed decision-making, they are not a standalone solution for mitigating the risks posed by deepfakes and misinformation. Future work should explore how these labels can be combined with other interventions, such as media literacy education and platform policies, to enhance their effectiveness. Furthermore, as AI-generated content evolves, continuous refinement of label design and further empirical testing is necessary to ensure that these interventions remain relevant and impactful.

\begin{acks}
This research was supported by the \href{https://university.open.ac.uk/centres/protecting-women-online/}{Open University's Centre for Protecting Women Online}, funded by the Research England Expanding Excellence in England (E3) Programme. The authors would also like to thank Michael Bernstein from Stanford University HCI group for the valuable feedback on revising the paper based on reviewer comments. 
\end{acks}

\bibliographystyle{ACM-Reference-Format}
\bibliography{sample-base}

\appendix

\section{Experiment Materials  }

\subsection{Image Sources for the Experiment }
\label{sec:Image sources for the experiment}
The images as shown in the figure \ref{fig:DFandR} used in this survey were obtained from various sources online and we acknowledge the copyright of the original creators of this content. Our use of these images is covered by the copyright exception of criticism and review - in this context for the review and criticism of the labels for AI-generated or synthetically generated content. Our sources for these images are presented in the table below. Please note since each image had higher resolution which exceeds our MB capacity, we only added the \underline{hyperlinks} to the source file. We included each of the images in a supplementary file.
\begin{figure}
    \centering
    \includegraphics[width=1.0\linewidth]{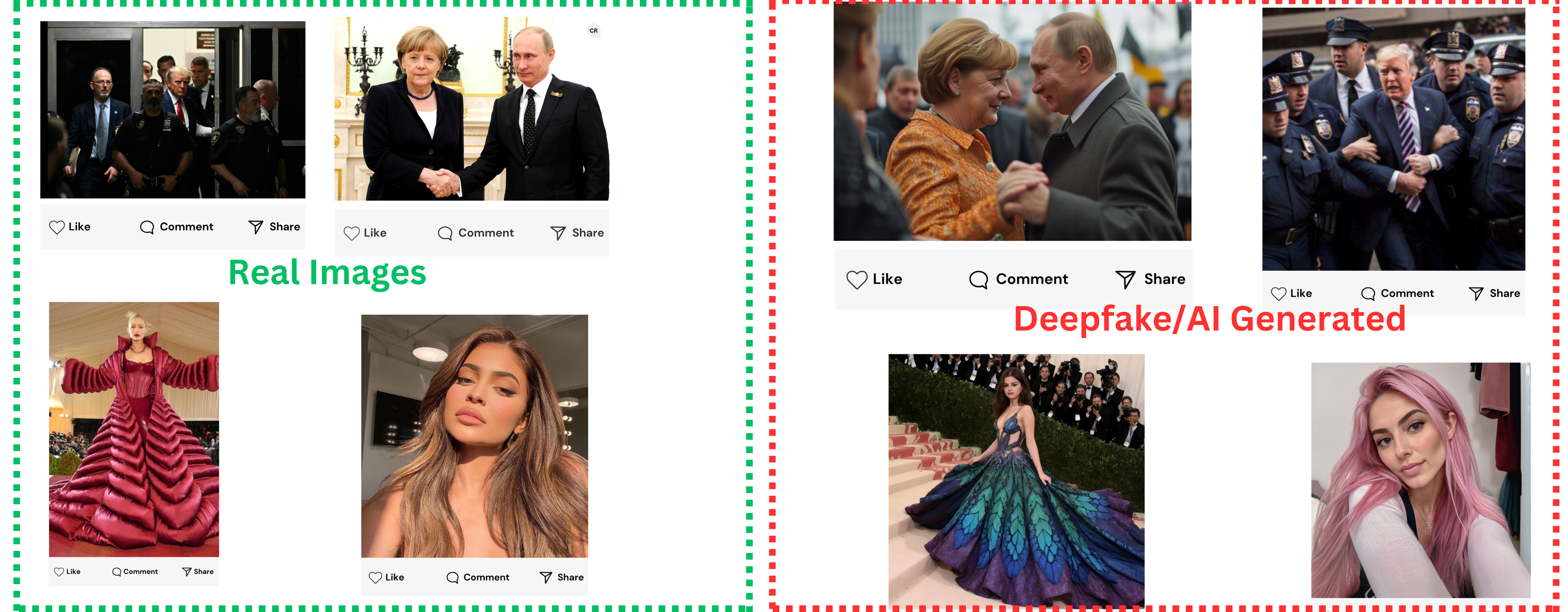}
    \caption{The images used to evaluate the user perceptions}
    \label{fig:DFandR}
\end{figure}

\begin{landscape}
\begin{longtable}{|p{4cm}|p{6cm}|p{6cm}|}
\caption{Summary of sources for AI-generated and real images used in the experiment with hyperlinks to the original files.} \label{table:ImageSources} \\
\hline
\textbf{Image Subject} & \textbf{AI Generated} & \textbf{Real} \\
\hline
\textbf{Politics 1} & 
\textbf{Image P1a:} Police officers surround and hold Donald Trump. 
\newline Source: \href{https://jgrj.law.uiowa.edu/news/2023/04/deepfake-technology-poses-threat-reality}{Deepfake Technology Poses Threat to Reality} & 
\textbf{Image P1b:} Police officers in front of Donald Trump as he leaves court. 
\newline Source: \href{https://www.msnbc.com/alex-wagner-tonight/watch/overwhelming-evidence-against-trump-dulls-benefits-of-supreme-court-gift-ruling-in-immunity-case-214199877974}{MSNBC: Overwhelming Evidence Against Trump} \\
\hline
\textbf{Politics 2} & 
\textbf{Image P2a:} Angela Merkel and Vladimir Putin in a dancing pose. 
\newline Source: \href{https://www.meinbezirk.at/wien/imagepost/deepfake-beispiel-putin-und-merkel-beim-walzer-tanzen_i661048}{Deepfake Example - Putin and Merkel Dancing} 
& 
\textbf{Image P2b:} Angela Merkel and Vladimir Putin shaking hands. 
\newline Source: \href{https://commons.wikimedia.org/wiki/File:Vladimir_Putin_and_Angela_Merkel_May_2015.jpg}{Wikimedia - Putin and Merkel May 2015} \\
\hline
\textbf{Entertainment 1} & 
\textbf{Image E1a:} Celebrity in a blue-purple gown on the staircase with photographers in the background. 
\newline Source: \href{https://www.facebook.com/photo/?fbid=122141558768108325&set=a.122108530472108325}{Facebook - Celebrity in Blue-Purple Gown} 
& 
\textbf{Image E1b:} Celebrity in a red gown on the red carpet. 
\newline Source: \href{https://www.thecut.com/2022/05/the-best-dressed-celebrities-on-the-2022-met-gala-red-carpet.html}{The Cut - Best Dressed Celebrities at 2022 Met Gala} \\
\hline
\textbf{Entertainment 2} & 
\textbf{Image E2a:} Selfie of celebrity with long pink hair. 
\newline Source: \href{https://www.facebook.com/photo/?fbid=10160800515835709&set=a.10150788826945709}{Facebook - Celebrity with Pink Hair}
& 
\textbf{Image E2b:} Selfie of celebrity with brown hair. 
\newline Source: \href{https://achcollection.com/trends/lifestyle/top-12-instagram-influencers-in-the-usa-you-must-know-about/}{ACH Collection - Top 12 Instagram Influencers in the USA} \\
\hline
\end{longtable}
\end{landscape}

\section{Detail Demographics of Data }
\label{sec: Demographicdata}


\label{sec:SocialMplatforms}
\begin{figure*}
    \centering
    \includegraphics[width=1.0\linewidth]{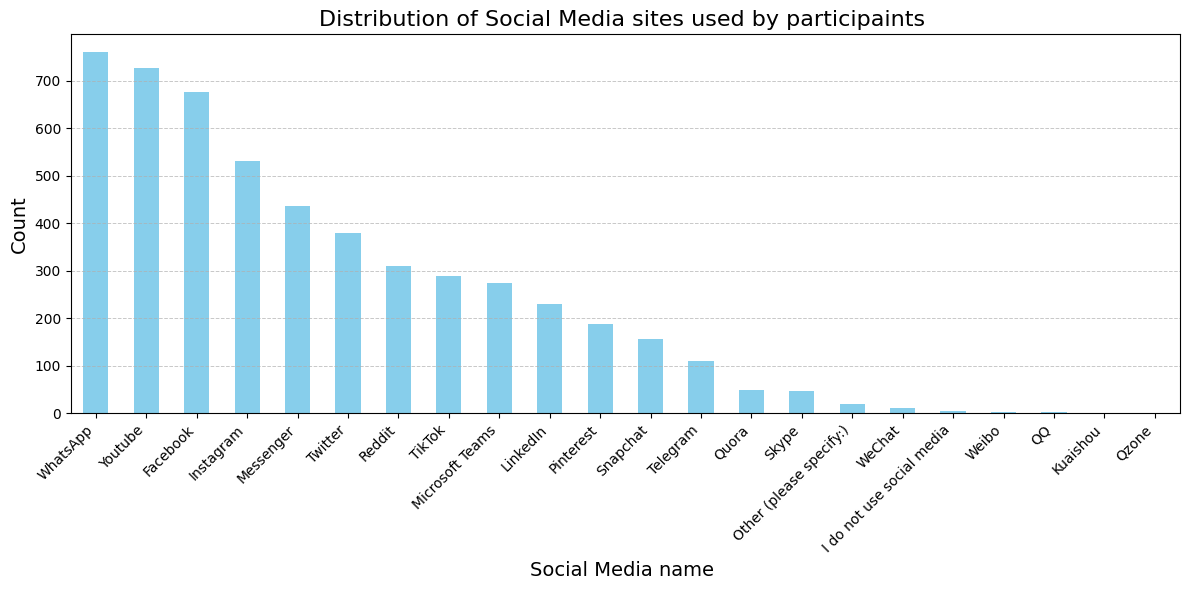}
    \caption{Types of social media use by the participants with the frequency of user}
    \label{fig:socialmeida}
\end{figure*}

\label{sec:age_Demographics}
\begin{table}[h!]
\centering
\caption{ Summary of Post\_Q1 Response on the age group Categories with Counts and Percentages}
\begin{tabular}{lcc}
\toprule
\textbf{Response Category} & \textbf{Count} & \textbf{Percentage (\%)} \\
\midrule
51-61 & 248 & 27.16 \\
29-39 & 179 & 19.61 \\
40-50 & 170 & 18.62 \\
18-28 & 151 & 16.54 \\
62-72 & 140 & 15.33 \\
Other age (please specify) &	23 	& 2.52\\
\bottomrule
\end{tabular}
\end{table}
}

\label{sec:Pronouns_Demographics}
\begin{table}[h!]
\centering
\caption{Summary of Post\_Q2 Responses on Pronouns}
\begin{tabular}{l r r}
\toprule
\textbf{Category} & \textbf{Count} & \textbf{Percentage (\%)} \\ 
\midrule
She/Her                   & 459 & 50.38 \\ 
He/Him                    & 418 & 45.88 \\ 
I prefer not to answer     & 30  & 3.29  \\ 
No Response (NaN)          & 2   & 0.22  \\ 
They/Them                  & 2   & 0.22  \\ 
\bottomrule
\end{tabular}
\label{tab:post_q2_distribution}
\end{table}


\label{sec:Q1}
\begin{table}[h!]
\caption{Q1: How many hours do you browse the Internet in a day?}
\centering
\begin{tabular}{l r r}
\toprule
\textbf{Category} & \textbf{Count} & \textbf{Percentage (\%)} \\ 
\midrule
2-3 hours a day    & 248 & 27.16 \\ 
1-2 hours per day  & 231 & 25.30 \\ 
3-4 hours a day    & 196 & 21.47 \\ 
5+ hours a day     & 172 & 18.84 \\ 
Less than 1 hour   & 65  & 7.12  \\ 
\bottomrule
\end{tabular}
\label{tab:q1_hours_internet}
\end{table}

\label{sec:Q4}
\begin{table}
\centering
\caption{Q4: How likely are you to use the social media platform to type messages, and upload content such as videos, photos, or voice/music on social media?}
\begin{tabular}{lrr}
\toprule
\textbf{Category} & \textbf{Count} & \textbf{Percentage (\%)} \\ 
\midrule
Somewhat likely              & 384 & 42.15 \\ 
Extremely likely             & 227 & 24.92 \\ 
Somewhat unlikely            & 139 & 15.26 \\ 
Neither likely nor unlikely  & 81  & 8.89  \\ 
Extremely unlikely           & 78  & 8.56  \\ 
No Response (NaN)            & 2   & 0.22  \\ 
\bottomrule
\end{tabular}
\label{tab:q4_social_media_usage}
\end{table}

\label{sec:Q5}
\begin{table}
\centering
\caption{Q5: How likely are you to engage (like, comment, other reactions) with such content posted by others on social media?}
\begin{tabular}{lrr}
\toprule
\textbf{Category} & \textbf{Count} & \textbf{Percentage (\%)} \\ 
\midrule
Somewhat likely              & 390 & 42.81 \\ 
Extremely likely             & 229 & 25.14 \\ 
Neither likely nor unlikely  & 115 & 12.62 \\ 
Somewhat unlikely            & 112 & 12.29 \\ 
Extremely unlikely           & 62  & 6.81  \\ 
No Response (NaN)            & 3   & 0.33  \\ 
\bottomrule
\end{tabular}
\label{tab:q4_social_media_likelihood}
\end{table}

\label{sec:Q2}
\begin{table}[h!]
\centering
\caption{Q2: How many hours do you use social media platforms in a day?}
\begin{tabular}{lrr}
\toprule
\textbf{Category} & \textbf{Count} & \textbf{Percentage (\%)} \\ 
\midrule
1-2 hours per day    & 348 & 38.20 \\ 
Less than 1 hour     & 219 & 24.04 \\ 
2-3 hours a day      & 183 & 20.09 \\ 
3-4 hours per day    & 98  & 10.76 \\ 
5+ hours per day     & 63  & 6.92  \\ 
\bottomrule
\end{tabular}
\label{tab:q2_hours_browse}
\end{table}


\begin{landscape}
\section{Analysis Tables Social Media Engagements    }
\subsection{Q3 Does having a label affect the engagement with the images?  }
\label{sec:Q3 }
\begin{tabular}{lrrrrrrrrrrr}
\toprule
 Group &  F\_statistic &      p\_value &  control\_mean &  treatment\_mean &  control\_std &  treatment\_std &  control\_size &  treatment\_size &  df\_between &  df\_within &  eta\_squared \\
\midrule
 TRT1\_ &    18.322240 & 1.997664e-05 &      2.904321 &        3.216954 &     1.359732 &       1.315382 &           648 &             696 &           1 &       1342 &     0.025461 \\
 TRT2\_ &    16.816412 & 4.369561e-05 &      2.904321 &        3.206845 &     1.359732 &       1.318509 &           648 &             672 &           1 &       1318 &     0.025461 \\
 TRT3\_ &    25.039039 & 6.377015e-07 &      2.904321 &        3.287202 &     1.359732 &       1.416091 &           648 &             672 &           1 &       1318 &     0.025461 \\
 TRT4\_ &    22.510440 & 2.320552e-06 &      2.904321 &        3.250000 &     1.359732 &       1.268161 &           648 &             656 &           1 &       1302 &     0.025461 \\
 TRT5\_ &    33.799364 & 7.695506e-09 &      2.904321 &        3.339564 &     1.359732 &       1.326682 &           648 &             642 &           1 &       1288 &     0.025461 \\
 TRT6\_ &    57.080250 & 7.849279e-14 &      2.904321 &        3.460366 &     1.359732 &       1.295495 &           648 &             656 &           1 &       1302 &     0.025461 \\
 TRT7\_ &    19.816860 & 9.256332e-06 &      2.904321 &        3.233025 &     1.359732 &       1.295656 &           648 &             648 &           1 &       1294 &     0.025461 \\
 TRT8\_ &    39.784285 & 3.849147e-10 &      2.904321 &        3.356105 &     1.359732 &       1.256237 &           648 &             688 &           1 &       1334 &     0.025461 \\
 TRT9\_ &    65.965428 & 1.036441e-15 &      2.904321 &        3.500000 &     1.359732 &       1.316801 &           648 &             686 &           1 &       1332 &     0.025461 \\
TRT10\_ &    19.631759 & 1.032084e-05 &      2.904321 &        3.259740 &     1.359732 &       1.252557 &           648 &             462 &           1 &       1108 &     0.025461 \\
\bottomrule
\end{tabular}

\subsection{Q4 Likelihood to  "Like" react?  }
\label{sec:Q4 Likelihood to  "Like" react? }
\begin{table}[ht]
\centering
\begin{tabular}{llllllllllll}
\toprule
Group  & F\_statistic & p\_value  & control\_mean & treatment\_mean & control\_std & treatment\_std & control\_size & treatment\_size & df\_between & df\_within & eta\_squared \\
\midrule
TRT1\_  & 9.068800  & 0.002649  & 1.666667 & 1.510057 & 0.990697 & 0.914388 & 648 & 696 & 1 & 1342 & 0.002679 \\
TRT2\_  & 0.091626  & 0.762167  & 1.666667 & 1.683036 & 0.990697 & 0.972453 & 648 & 672 & 1 & 1318 & 0.002679 \\
TRT3\_  & 3.915777  & 0.048043  & 1.666667 & 1.561012 & 0.990697 & 0.947662 & 648 & 672 & 1 & 1318 & 0.002679 \\
TRT4\_  & 0.037721  & 0.846036  & 1.666667 & 1.655488 & 0.990697 & 1.083510 & 648 & 656 & 1 & 1302 & 0.002679 \\
TRT5\_  & 0.035183  & 0.851241  & 1.666667 & 1.677083 & 0.990697 & 1.024206 & 648 & 672 & 1 & 1318 & 0.002679 \\
TRT6\_  & 0.000118  & 0.991727  & 1.666667 & 1.667963 & 0.990697 & 0.994058 & 648 & 656 & 1 & 1302 & 0.002679 \\
TRT7\_  & 0.505065  & 0.477699  & 1.666667 & 1.629104 & 0.990697 & 0.965931 & 648 & 648 & 1 & 1294 & 0.002679 \\
TRT8\_  & 0.000448  & 0.983181  & 1.666667 & 1.664844 & 0.990697 & 1.026517 & 648 & 688 & 1 & 1334 & 0.002679 \\
TRT9\_  & 0.046239  & 0.829737  & 1.666667 & 1.647063 & 0.990697 & 1.019539 & 648 & 784 & 1 & 1430 & 0.002679 \\
TRT10\_ & 1.682592  & 0.194789  & 1.666667 & 1.596842 & 0.990697 & 0.949440 & 648 & 475 & 1 & 1121 & 0.002679 \\
\bottomrule
\end{tabular}
\caption{ANOVA Results for Q4 with Eta Squared (10 groups)}
\end{table}
\end{landscape}

\section{Effect Size of the Believability in Treatment Groups}
\label{sec:beliveability effect size}
\begin{table*} [h!]
\centering
\begin{tabular}{ll}
\toprule
\textbf{Warning label Group} & \textbf{Effect Size Cohen's d} \\
\midrule
TRT9\_  & 0.445259 \\
TRT6\_  & 0.418770 \\
TRT8\_  & 0.345544 \\
TRT5\_  & 0.323990 \\
TRT3\_  & 0.275710 \\
TRT10\_ & 0.270034 \\
TRT4\_  & 0.262981 \\
TRT7\_  & 0.247503 \\
TRT1\_  & 0.233841 \\
TRT2\_  & 0.225949 \\ 
\bottomrule
\end{tabular}
\caption{Believability Effect sizes by each treatment group.}
\label{tab:effect_sizes}
\end{table*}

\begin{landscape}

\subsection{Q5 Likelihood to  "Comment"?  }
\label{sec:Q5 Likelihood to  "Comment" }
\begin{table}[ht]
\centering
\begin{tabular}{llllllllllll}
\toprule
Group  & F\_statistic & p\_value  & control\_mean & treatment\_mean & control\_std & treatment\_std & control\_size & treatment\_size & df\_between & df\_within & eta\_squared \\
\midrule
TRT1\_  & 0.660804  & 0.416420  & 1.450617 & 1.415230 & 0.780175 & 0.812084 & 648 & 696 & 1 & 1342 & 0.001794 \\
TRT2\_  & 0.279897  & 0.596858  & 1.450617 & 1.428571 & 0.780175 & 0.732506 & 648 & 672 & 1 & 1318 & 0.001794 \\
TRT3\_  & 1.288541  & 0.256525  & 1.450617 & 1.401515 & 0.780175 & 0.782961 & 648 & 660 & 1 & 1306 & 0.001794 \\
TRT4\_  & 2.675141  & 0.102168  & 1.450617 & 1.528963 & 0.780175 & 0.939844 & 648 & 656 & 1 & 1302 & 0.001794 \\
TRT5\_  & 2.958991  & 0.085641  & 1.450617 & 1.532508 & 0.780175 & 0.925070 & 648 & 646 & 1 & 1292 & 0.001794 \\
TRT6\_  & 0.002859  & 0.957367  & 1.450617 & 1.448171 & 0.780175 & 0.867993 & 648 & 656 & 1 & 1302 & 0.001794 \\
TRT7\_  & 2.412928  & 0.120582  & 1.450617 & 1.523148 & 0.780175 & 0.895507 & 648 & 648 & 1 & 1294 & 0.001794 \\
TRT8\_  & 0.001085  & 0.973733  & 1.450617 & 1.452035 & 0.780175 & 0.790953 & 648 & 688 & 1 & 1334 & 0.001794 \\
TRT9\_  & 0.117589  & 0.731715  & 1.450617 & 1.465561 & 0.780175 & 0.851971 & 648 & 784 & 1 & 1430 & 0.001794 \\
TRT10\_ & 1.019855  & 0.312771  & 1.450617 & 1.404211 & 0.780175 & 0.731820 & 648 & 475 & 1 & 1121 & 0.001794 \\
\bottomrule
\end{tabular}
\caption{ANOVA Results for Q5 with Eta Squared (10 groups)}
\end{table}

\subsection{Q6 Likelihood to  "Share"?  }
\label{sec:Q7 Likelihood to  "Share" }

\begin{table}[ht]
\centering
\begin{tabular}{llllllllllll}
\toprule
Group  & F\_statistic & p\_value  & control\_mean & treatment\_mean & control\_std & treatment\_std & control\_size & treatment\_size & df\_between & df\_within & eta\_squared \\
\midrule
TRT1\_  & 0.987507  & 0.320532  & 2.035494 & 2.117816 & 1.453064 & 1.573094 & 648 & 696 & 1 & 1342 & 0.00353 \\
TRT2\_  & 0.685234  & 0.407940  & 2.035494 & 2.102679 & 1.453064 & 1.492000 & 648 & 672 & 1 & 1318 & 0.00353 \\
TRT3\_  & 0.000007  & 0.997860  & 2.035494 & 2.035714 & 1.453064 & 1.527108 & 648 & 672 & 1 & 1318 & 0.00353 \\
TRT4\_  & 0.752072  & 0.385983  & 2.035494 & 2.107477 & 1.453064 & 1.525272 & 648 & 642 & 1 & 1288 & 0.00353 \\
TRT5\_  & 3.850970  & 0.049927  & 2.035494 & 2.196429 & 1.453064 & 1.521704 & 648 & 672 & 1 & 1318 & 0.00353 \\
TRT6\_  & 2.540184  & 0.111231  & 2.035494 & 2.152558 & 1.453064 & 1.518747 & 648 & 656 & 1 & 1302 & 0.00353 \\
TRT7\_  & 0.007372  & 0.931327  & 2.035494 & 2.040394 & 1.453064 & 1.494687 & 648 & 648 & 1 & 1294 & 0.00353 \\
TRT8\_  & 0.251194  & 0.616476  & 2.035494 & 2.075799 & 1.453064 & 1.538432 & 648 & 688 & 1 & 1334 & 0.00353 \\
TRT9\_  & 1.065351  & 0.302130  & 2.035494 & 2.124133 & 1.453064 & 1.497918 & 648 & 784 & 1 & 1430 & 0.00353 \\
TRT10\_ & 1.415370  & 0.234478  & 2.035494 & 2.101053 & 1.453064 & 1.530255 & 648 & 475 & 1 & 1121 & 0.00353 \\
\bottomrule
\end{tabular}
\caption{ANOVA Results for Q6 with Eta Squared (10 groups)}
\end{table}
\end{landscape}

\label{sec:Posthoc Test in Q4 Likelihood to like}
\begin{table*}
\begin{tabular}{llrrrrl}
\textbf{Group 1} & \textbf{Group 2} & \textbf{Mean Diff} & \textbf{p-adj} & \textbf{Reject} & \textbf{Cohen's d} \\
\midrule
\textsc{Entertainment}\(_{Control}\) & \textsc{Entertainment}\(_{Treatment}\) & -0.0182 & 0.9901 & False & -0.0178 \\
\textsc{Entertainment}\(_{Control}\) & \textsc{Political}\(_{Control}\)        & 0.0679  & 0.8323 & False & 0.0665  \\
\textsc{Entertainment}\(_{Control}\) & \textsc{Political}\(_{Treatment}\)      & 0.0639  & 0.7049 & False & 0.0626  \\
\textsc{Entertainment}\(_{Treatment}\) & \textsc{Political}\(_{Control}\)      & 0.0861  & 0.4695 & False & 0.0843  \\
\textsc{Entertainment}\(_{Treatment}\) & \textsc{Political}\(_{Treatment}\)    & 0.0821  & 0.0061 & True  & 0.0804  \\
\textsc{Political}\(_{Control}\)       & \textsc{Political}\(_{Treatment}\)    & -0.0040 & 0.9999 & False & -0.0039 \\
\bottomrule
\end{tabular}    
\caption{ Posthoc Test in Q4 Likelihood to Like   }
\end{table*}

\label{sec:Posthoc Test in Q5 Likelihood to Comment}
\begin{table*}
\begin{tabular}{lllllll}
\toprule
\textbf{Group 1} & \textbf{Group 2} & \textbf{Mean Diff} & \textbf{p-adj} & \textbf{Reject} & \textbf{Cohen's d} \\
\midrule
\textsc{Entertainment}\(_{Control}\) & \textsc{Entertainment}\(_{Treatment}\) & -0.0362 & 0.8776 & False & -0.0435 \\
\textsc{Entertainment}\(_{Control}\) & \textsc{Political}\(_{Control}\)        & 0.0494  & 0.8743 & False & 0.0594  \\
\textsc{Entertainment}\(_{Control}\) & \textsc{Political}\(_{Treatment}\)      & 0.1061  & 0.1261 & False & 0.1275  \\
\textsc{Entertainment}\(_{Treatment}\) & \textsc{Political}\(_{Control}\)      & 0.0856  & 0.2893 & False & 0.1029  \\
\textsc{Entertainment}\(_{Treatment}\) & \textsc{Political}\(_{Treatment}\)    & 0.1423  & 0.0000 & True  & 0.1711  \\
\textsc{Political}\(_{Control}\)       & \textsc{Political}\(_{Treatment}\)    & 0.0567  & 0.6458 & False & 0.0682  \\
\bottomrule
\end{tabular}
\caption{ Posthoc Test in Q5 Likelihood to Comment }
\end{table*}

\label{sec:Posthoc Test in Q6 Likelihood to share}
\begin{table*}
\begin{tabular}{lllllll}
\toprule
\textbf{Group 1} & \textbf{Group 2} & \textbf{Mean Diff} & \textbf{p-adj} & \textbf{Reject} & \textbf{Cohen's d} \\
\midrule
\textsc{Entertainment}\(_{Control}\) & \textsc{Entertainment}\(_{Treatment}\) & 0.1968  & 0.1150 & False & 0.1299  \\
\textsc{Entertainment}\(_{Control}\) & \textsc{Political}\(_{Control}\)        & 0.3796  & 0.0078 & True  & 0.2505  \\
\textsc{Entertainment}\(_{Control}\) & \textsc{Political}\(_{Treatment}\)      & 0.3315  & 0.0010 & True  & 0.2187  \\
\textsc{Entertainment}\(_{Treatment}\) & \textsc{Political}\(_{Control}\)      & 0.1828  & 0.1621 & False & 0.1206  \\
\textsc{Entertainment}\(_{Treatment}\) & \textsc{Political}\(_{Treatment}\)    & 0.1347  & 0.0017 & True  & 0.0889  \\
\textsc{Political}\(_{Control}\)       & \textsc{Political}\(_{Treatment}\)    & -0.0482 & 0.9477 & False & -0.0318 \\
\bottomrule
\end{tabular}
\caption{Posthoc Test in Q6 Likelihood to Share}
\end{table*}

\label{sec:Q10 seeing}

\begin{table*}[ht]
\centering
\begin{tabular}{lcccc}
\toprule
Source     & Sum of Squares & df & F-Statistic & p-value (PR(>F)) \\
\midrule
TRT\_Group & 24.469403      & 9  & 1.892228    & 0.048451         \\
Residual   & 9337.999179    & 6499 & -          & -                \\
\bottomrule
\end{tabular}
\caption{ANOVA Results for TRT Groups (Q10)}
\end{table*}

\begin{table*}[ht]
\centering
\begin{tabular}{lcccccc}
\toprule
Group 1 & Group 2 & Mean Difference & p-adj & Lower Bound & Upper Bound & Reject \\
\midrule
TRT2\_  & TRT6\_  & -0.2192         & 0.0297 & -0.4274     & -0.011      & True   \\
\bottomrule
\end{tabular}
\caption{Tukey's HSD post-hoc Results (Significant Comparisons)}
\end{table*}

\begin{table*}[ht]
\begin{tabular}{lcccccc}
\toprule
TRT Group & F-statistic & p-value  & Eta Squared & Sample Size & Mean Value & Std Value \\
\midrule
TRT1      & 1.309946    & 0.242615 & 0.013153   & 87.0        & 3.429598   & 1.242116  \\
TRT2      & 2.502549    & 0.015241 & 0.025704   & 84.0        & 3.575893   & 1.116119  \\
TRT3      & 1.458434    & 0.179175 & 0.015142   & 84.0        & 3.500000   & 1.176018  \\
TRT4      & 0.748098    & 0.631221 & 0.008017   & 82.0        & 3.403963   & 1.265340  \\
TRT5      & 2.431512    & 0.018260 & 0.024993   & 84.0        & 3.441964   & 1.173341  \\
TRT6      & 1.654390    & 0.136897 & 0.017189   & 82.0        & 3.353659   & 1.150789  \\
TRT7      & 0.986178    & 0.424198 & 0.010082   & 83.0        & 3.307229   & 1.135978  \\
TRT8      & 1.725132    & 0.111472 & 0.017867   & 84.0        & 3.530580   & 1.222098  \\
TRT9      & 0.976319    & 0.418210 & 0.010000   & 85.0        & 3.411765   & 1.207199  \\
TRT10     & 1.268769    & 0.276423 & 0.012784   & 84.0        & 3.349206   & 1.176446  \\
\bottomrule
\end{tabular}
\caption{ANOVA Results for TRT Groups (Q10)}
\end{table*}

\label{sec: Trusttolabelandplatform}
\begin{figure*}
    \centering
    \includegraphics[width=0.75\linewidth]{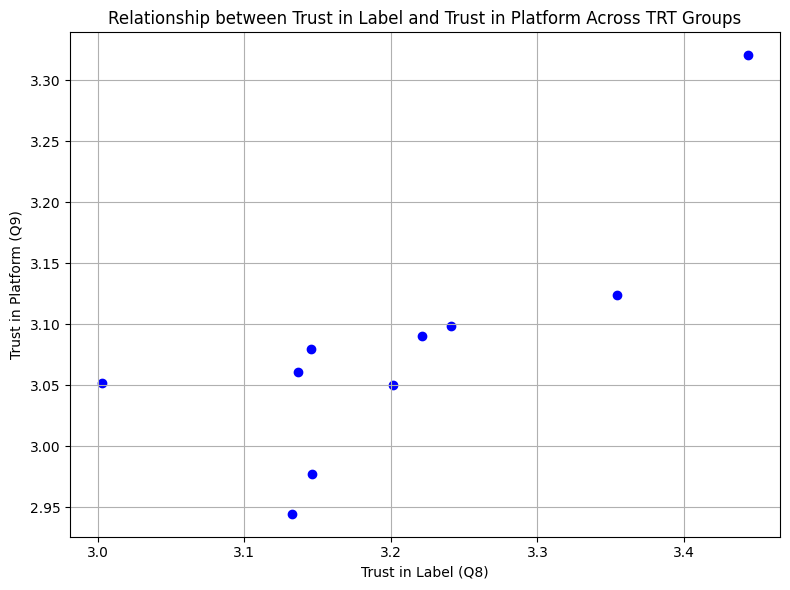}
    \caption{Relationship Between warning Label trust and trust to the Platform showed a significant correlation  \(r(85)=.73, p < .001\)}
    \label{fig:relationshipBtwnLabelandPlatform}
    \Description{ }
\end{figure*}

\end{document}